\documentclass[useAMS, usenatbib, usegraphicx]{mn2e} 
\usepackage{deluxetable} 
\usepackage{amsmath}

\title{Parker Winds Revisited: An Extension to Disc Winds} 
\author[Waters \& Proga] 
{Timothy R. Waters$^1$, Daniel Proga$^{1,2,}$\thanks{Permanent address: UNLV Mailstop 4002, 4505 South Maryland Parkway, Las Vegas, NV, 89154; dproga@physics.unlv.edu} 
\\
$^1$Department of Physics \& Astronomy, University of Nevada, Las Vegas,
NV 89154; waterst3@unlv.nevada.edu \\
$^2$Princeton University Observatory, Peyton Hall, Princeton, NJ 08544; dproga@astro.princeton.edu}  

\numberwithin{equation}{section}

\newcommand{\mnras}{MNRAS} 
\newcommand{\apj}{ApJ} 
\newcommand{\apjs}{ApJ, Suppl.}
\newcommand{\apjl}{ApJ, Lttr.}
\newcommand{\aap}{A\&A}
\newcommand{\icarus}{ICARUS}
\newcommand{\ssr}{Space Sci. Rev.}
\newcommand{\araa}{ARA\&A}
\newcommand{\apss}{Astrophysics \& Space Sci.}
\newcommand{\pasj}{Pub. ASJ}
\newcommand{\planss}{Planetary Space Sci.}

\newcommand{\beq}{\begin{equation}}
\newcommand{\seq}{\end{equation}}

\newcommand{\fig}{$\text{Figure }$\ref}
\newcommand{\tbl}[1]{$\text{Table #1}$}
\newcommand{\apx}{$\text{Appendix }$\ref}
\renewcommand{\sec}[1]{\S{\ref{#1}}}

\renewcommand{\v}[1]{\ensuremath{\mathbf{#1}}} % for vectors
\newcommand{\gv}[1]{\ensuremath{\mbox{\boldmath$ #1 $}}} % for vectors of Greek letters
\newcommand{\avg}[1]{\left< #1 \right>} % for average
\renewcommand{\d}[2]{\frac{d #1}{d #2}} % for derivatives
\newcommand{\pd}[2]{\frac{\partial #1}{\partial #2}} 
\renewcommand{\div}[1]{\gv{\nabla} \cdot #1} % for divergence
\newcommand{\f}{\frac}  %shorter fraction
\newcommand{\eqn}[1]{equation (\ref{#1})} %call equations
\newcommand{\bigb}[1]{\left[ #1 \right]} %big brackets
\newcommand{\bigp}[1]{\left( #1 \right)} %big parenthesis

\newcommand{\x}{\chi}
\newcommand{\p}{\rho}
\newcommand{\rh}{\hat{r}}

\newcommand{\lp}{\lambda_o}
\newcommand{\lc}{\lambda_c}

\newcommand{\gfac}{\frac{1}{\gamma -1}}
\newcommand{\gfacb}{\frac{\gamma+1}{\gamma -1}}

\newcommand{\igfacb}{\frac{\gamma-1}{\gamma +1}}
\newcommand{\gmone}{\gamma -1}
\newcommand{\gpone}{\gamma +1}

\begin{document} 

\maketitle

\begin{abstract} 
A simple 1D dynamical model of thermally driven disc winds is proposed, based on the results of recent, 2.5D axi-symmetric simulations.   Our formulation of the disc wind problem is in the spirit of the original Parker (1958) and Bondi (1952) problems, namely we assume an elementary flow configuration consisting of an outflow following pre-defined trajectories in the presence of a central gravitating point mass.  Viscosity and heat conduction are neglected.  We consider two different streamline geometries, both comprised of straight lines in the $(x,z)$-plane: (i) streamlines that converge to a geometric point located at $(x,z)=(0,-d)$ and (ii) streamlines that emerge at a constant inclination angle from the disc midplane (the $x$-axis, as we consider geometrically thin accretion discs).  The former geometry is commonly used in kinematic models to compute synthetic spectra, while the latter, which exhibits self-similarity, is likely unused for this purpose, although it easily can be with existing kinematic models.  We make the case that it should be, i.e. that geometry (ii) leads to transonic wind solutions with substantially different properties owing to its lack of streamline divergence.  Both geometries can be used to complement recent efforts to estimate photoevaporative mass loss rates from protoplanetary discs.  Pertinent to understanding our disc wind results, which are also applicable to X-ray binaries and active galactic nuclei, is a focused discussion on lesser known properties of classic Parker wind solutions.  We find that the parameter space corresponding to decelerating Parker wind solutions is made larger due to rotation and leads instead to disc wind solutions that always accelerate after the bulk velocity is slowed to a minimum value.  Surprisingly, Keplerian rotation may allow for two different transonic wind solutions for the same physical conditions.        
\end{abstract} 


\section{Introduction} 
The classic Parker model has served as a paradigm wind solution for over half a century now.  First developed for the Sun as a model of the solar wind, it shows the essential features of one-dimensional (1D), steady state wind models, namely that transonic solutions typically involve a transition through a critical point and have an X-type solution topology.  The analytic solutions to both the original isothermal (with adiabatic index $\gamma=1$) Parker wind model (Parker, 1958) and its polytropic fluid $(1<\gamma<5/3)$ extension (Parker, 1960) serve a dual purpose.  On the one hand, they are beneficial for obtaining insight into the theory of outflows in general, as well as for gaining intuition into the subtleties that arise when solving wind equations analytically (for an in depth perspective see K\"onigl \& Salmeron 2011 and references therein).  On the other hand, Parker wind solutions have proven useful for assessing the accuracy of numerical simulations (e.g., Keppens \& Goedbloed 1999; Font et al. 2004; Tian et al. 2005; Stone \& Proga 2009).  To that end, one goal of this paper is to present a simplified dimensionless formulation of Parker winds and to provide formulae commonly used for numerical testing purposes.  At the same time, we address certain aspects of the polytropic Parker problem that have been a source of confusion in the literature.  Specifically, we clarify the properties of spherically symmetric Parker winds in the range $3/2<\gamma<5/3$ and the corresponding range of $\gamma$ when angular momentum is added to the problem.  
  
The primary focus of this paper is to present solutions to Parker-like winds emanating as a biconical flow, the geometry commonly used to model accretion disc winds.  Parker winds have been instrumental in uncovering other physical processes that can drive winds in stars, and led to the development of both line-driven (Castor et al. 1975, hereafter CAK) and magneto-centrifugally driven (beginning with the solution of Weber \& Davis 1967) wind theory.  The current state of the art in stellar wind theory owes much of its development to the systematic assessment of how the inclusion of various physical terms and geometrical effects in the hydrodynamic equations alters the solutions of Parker winds.  Studies of disc winds stand to benefit from rigorously repeating this procedure using counterpart, 1D analytical disc wind models.   

Developing concrete baseline models analogous to Parker winds has proven to be a difficult task.  A major roadblock has been the uncertainty in the streamline geometry, i.e. the actual trajectory traversed by gas flowing out from the disc, as well as in the gravitational potential along these streamlines.  Another obvious and related difficulty is posed by the fact that accretion discs span many more orders of magnitude in physical size than do stars, and they can host radically different, spatially and temporally variable, thermodynamic environments.  Indeed, the outer radius of an accretion disc ranges from parsec scales for active galactic nuclei (AGN) down to within 1 AU for some circumstellar discs and the diverse physical conditions permit anything from infrequent outbursts to highly relativistic, steady jets.  It should come as no surprise then, that despite clear observational evidence of outflows from many systems, identifying the actual driving mechanisms, as well as determining the wind geometry, remains a challenge. 

Studies of disc winds therefore rely heavily on kinematic models in order to quickly explore the parameter space without assuming a particular driving mechanism.  For example, kinematic models have been employed to produce synthetic spectra for cataclysmic variables (CVs), systems in which even key properties such as the geometry, ionization structure, and mass-loss rates remain difficult to constrain (e.g., Noebauer et al. 2010 and references therein).  Early kinematic models assumed spherically symmetric outflows for simplicity (Drew \& Verbunt 1985; Mauche \& Raymond 1987).  The consensus picture of a biconical mass outflow originating from the inner disc was born out of the observed characteristics of resonance lines in CVs (C\'{o}rdova \& Mason 1985, Drew 1987).  This geometry was developed into a robust kinematic model by Shlosman \& Vitello (1993), who calculated the ionization structure of CV disc winds and solved a radiative transfer problem in lines using the Sobolev approximation.  Their kinematic model allowed for an arbitrary amount of streamline divergence.  

\setlength{\unitlength}{1.0in}
 \begin{figure}
 \centering
\includegraphics[scale = .35]{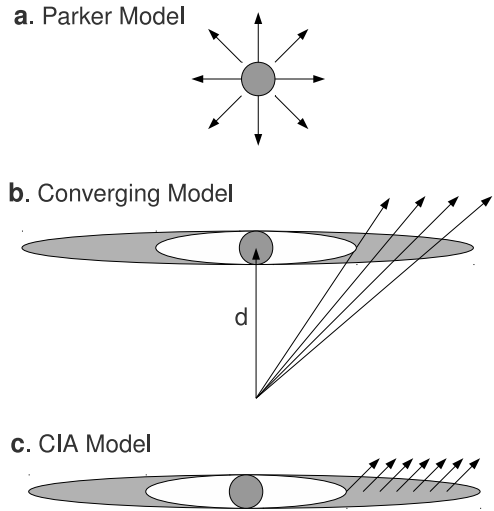}
\vspace{-.25in}
\caption{Pictorial representation of the streamline geometry addressed in this paper.  Neighboring streamlines diverge from each other in the (a) Parker and (b) Converging models, whereas in (c), the Constant Inclination Angle (CIA) model, there is no adjacent streamline divergence. } 
\label{modelsABC}
\end{figure}

Knigge et al. (1995) developed a different (Monte Carlo) code to solve the radiative transfer exactly.  Their choice of wind geometry is one instance of what we refer to as the Converging model, in that the streamline divergence is just such that all streamlines converge to a geometric point located a distance $d$ below the disc, as illustrated in \fig{modelsABC}b.  The Converging model, which has been called the ``displaced-dipole" model by others, has been used in conjunction with sophisticated radiative transfer simulations to model accretion disc spectra from massive young stellar objects (Sim et al. 2005),  active galactic nuclei (Sim et al. 2008), CV disc winds (Noebauer et. al 2010), classical T Tauri stars (Kurosawa et al. 2011), and young intermediate-mass Herbig Ae stars (Grinin \& Tambovtseva 2011).  Typically, these simulations use Monte Carlo procedures that can account for nearly all of the prominent resonance lines and thereby accurately calculate the ionization balance of the wind.  The Converging model has even been employed to calculate the neutron structure of neutrino-heated MHD disc winds associated with gamma-ray bursts (Metzger et al. 2008).   

In this paper we develop a simple \textit{dynamical} disc wind model that amounts to a generalization of the Parker model.  Rather than positing a velocity law as is done for kinematic models, the purpose of a dynamical model is to impose the physical conditions and solve for the wind velocity as a function of distance along a streamline.  This necessarily requires identifying a driving mechanism, i.e. a heating source in the case of thermally driven winds.  Much of the groundwork theory for the source of heating was laid down by Begelman et al. (1983, hereafter BMS83), who showed that Compton-heated coronae are qualitatively the same for both quasars and X-ray binaries.  That is, both galactic X-ray sources and the inner regions of AGN are expected to be heated via irradiation from a central X-ray source to high enough temperatures that thermal expansion alone gives rise to a disc wind.  

The hydrodynamic formulation of BMS83 established that it suffices to estimate 2D global wind properties with a 1D model that captures the essential physics.  Indeed, many predictions given by BMS83 were later confirmed by followup works that focused on the inherently two-dimensional radiative transfer problem (e.g., Ostriker et al. 1991, Woods et al. 1996, Proga \& Kallman 2002).  Of special interest here is the work by Woods et al. (1996), who added to the basic theory of BMS83 based on the outcome of their time-dependent, 2.5D simulations of thermally driven winds from AGN heated by Compton as well as non-Compton processes such as photoionization and line-cooling.  They provided an improved formula for the mass flux density as a function of disc radius and presented a detailed study of the flow topology and sonic surfaces for various spectral energy distributions.    

Both the results of Woods et al. (1996) and those of the more recent 2.5D time-dependent simulations of a thermally driven wind carried out by Luketic et al. (2010, see \fig{stefans} here) indicate that the streamline geometry is rather simple, displaying two distinct flow patterns.  Moreover, their results suggest that the Converging model may not be well-suited for sampling the entire wind, but rather only the inner portions of it.  The outer portion is better approximated by a model in which streamlines emerge at a constant inclination angle to the midplane (hence the name, the CIA Model -- see \fig{modelsABC}c).\footnote{The kinematic model used by Shlosman \& Vitello (1993) can accommodate CIA streamlines by setting $\theta_{min} = \theta_{max}$.}  It is our intention to study how this difference in geometry affects the hydrodynamics independent of the explicit heating mechanism taking place; we merely assume that the boundary of the flow (the disc midplane) has been heated to a high enough temperature to drive a thermal wind.  

\setlength{\unitlength}{1.0in}
 \begin{figure}
\begin{picture}(4,4)(-2.5,0)
\put(2,0){\includegraphics{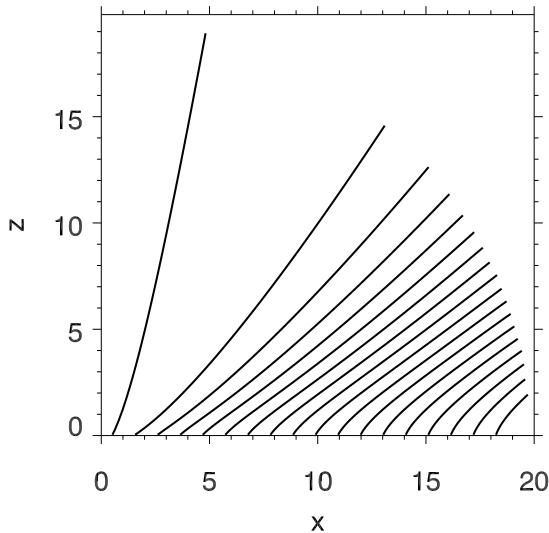}}
\end{picture}
\vspace{-.5 in}
\caption{Flow streamlines that resulted from the time-dependent, hydrodynamical simulation of a thermally driven wind (Luketic et al. 2010).  The $z$-axis is the rotation axis, while the $x$-axis is the disc midplane.  Streamlines at $x \ga 4 R_{IC}$, where $R_{IC}$ is the Compton radius, are self-similar. This figure gave motivation for the CIA model. } 
\label{stefans}
\end{figure}

This paper is organized as follows.   Background theory pertaining to the hydrodynamic foundations of wind theory, sources of thermal driving, and our choice of streamline geometry is presented in \sec{PLDWs}.  Our formulation is given in \sec{formulation}.  We make a comparison with simulations and present our results in \sec{results}.  We discuss subtle aspects of the classic Parker problem that are relevant to disc winds and reveal the effects of adding rotation in \sec{discussion}.  Finally, we summarize our findings and open questions in \sec{conclusions}. 

\section{Parker-like Disc Winds} \label{PLDWs}  
Aside from stellar winds, spherical Parker winds have been utilized to model AGN winds (e.g., Everett \& Murray 2007) and protoplanetary disc winds (e.g., Adams et al. 2004, Gorti \& Hollenbach 2009, Owen et al. 2012).  They are also very useful for conceptualizing the atmospheric escape process from planets (e.g., Tucker et al. 2012 and references therein).  By relaxing the assumption of spherical symmetry, we call the outflow traversing an azimuthally symmetric streamline geometry a \textit{Parker-like disc wind}.  The latter retains the same notion of a classic Parker wind, which we review in \sec{hydrostatics} before discussing concepts specific to disc winds.  

\subsection{Static Atmospheres vs. Parker Winds}\label{hydrostatics}

Parker winds describe a highly idealized fluid phenomenon: the steady state, spherically symmetric, hydrodynamic escape of an ideal gas with zero viscosity.  With these simplifications, and further supposing a fluid composed of only one species of gas, the Eulerian equations become analytically tractable.  A complete isothermal Parker wind solution consists of the density $\rho$ and velocity $v$ as a function of radius $r$.  Polytropic Parker wind solutions also include a temperature profile $T(r)$.  The family of isothermal ($\gamma=1$) transonic Parker wind solutions depends solely on one parameter, which is often called the hydrodynamic energy parameter (HEP) and is defined as
\beq \lp  = \f{GM_* m_p \mu}{\gamma kT_o r_o}.\label{HEPdefn}\seq
Here, $G$ is the gravitational constant, $M_*$ is the central body's mass, $m_p$ the proton mass, $\mu$ is the mean molecular weight, $k$ is Boltzmann's constant, and $T_o$ is the temperature at the base of a streamline, i.e. at the boundary radius $r_o$.  A polytropic equation of state (EoS) introduces a second dependence on the adiabatic index $\gamma$.  (The choice to incorporate $\gamma$ into the HEP was not made in the early papers on the solar wind, but it makes the dimensionless equations less cluttered.) 

Physically, Parker wind solutions model atmospheric coronae in hydro\emph{dynamic} equilibrium just as a familiar barometric law, which yields the variation of density with radius, models an atmosphere in hydro\emph{static} equilibrium.  Indeed, an isothermal barometric law (albeit one that accounts for a varying gravitational potential) can be considered the trivial Parker wind solution with $v(r)=0$, $T(r)=T_o=constant$, and 
\beq \rho(r) = \rho_o\exp\bigb{\lp(r_o/r) -\lp} \label{introbarometriclaw},\seq
where $\rho_o$ is taken to be the known density at a reference radius $r_o$.
Equation (\ref{introbarometriclaw}) is most commonly recognized as the solution to the equation of hydrostatic equilibrium, 
\beq \f{dP}{dr} = -\f{GM_*\rho}{r^2} \label{staticequilb} \seq
for the special case of an ideal gas with a pressure $P = \rho kT_o/m_p \mu$.  However, \eqn{introbarometriclaw} is readily seen to be the density profile found by taking the limit of a Parker wind solution with a slowly expanding atmosphere ($v \rightarrow 0$) and a small mass-loss rate (corresponding to an everywhere subsonic solution -- see \sec{LambertWfctsoln} for more discussion). 
Recovering a static atmosphere from a slowly expanding one hints at a correspondence between thermal escape processes in kinetic theory and fluid dynamics.

\bigskip
\subsubsection{Hydrodynamic Escape}
Parker winds capture the simplest example of a more general thermal escape process characterized by hydrodynamic escape.  Modern, more realistic models can account for additional physical processes and non-spherical geometries, but the underlying hydrodynamic, thermal escape mechanism is effectively isolated by Parker winds.  

The thermal escape process from a static atmosphere (i.e. evaporation or Jeans escape) is governed by kinetic theory and sometimes referred to as hydrostatic escape to distinguish (and emphasize) its relation to hydrodynamic escape (e.g., Seager 2010, pg. 448).  A parameter common to both the kinetic theory and fluid dynamics approaches to deriving a barometric law for a static (i.e. slowly evaporating) atmosphere is what we call the thermal energy parameter (TEP), 
\beq \tau \equiv \f{|\Phi|}{c_s^2} =  \lambda_o\f{r_o/r}{T(r)/T_o},\label{TEPdef}\seq
where $\Phi = -GM_*/r$ is the gravitational potential energy per unit mass and $c_s=\sqrt{\gamma kT/\mu m_p}$ is the adiabatic speed of sound.  The TEP is by definition a measure of the thermal energy of gas at every location in a central gravitational field, and the second equality permits us to interpret the HEP as just the TEP evaluated at some reference level $r_o$, at which the temperature is $T_o$.  The magnitude of the HEP at this level, as well as the asymptotic value of the TEP, governs which approach, fluid or kinetic, better models the escape process (for concrete examples, see Tucker et al. 2012; see also Owen \& Jackson 2012).  

Transonic hydrodynamic escape is associated with the important property that the pressure tends to zero asymptotically (Parker 1958, 1960, 1965).  That static atmospheres can lack this property  in essence provided the physical basis for Parker's original transonic solar wind solution, as the hydrostatic conduction model of Chapman (1957), the model of the extended solar corona that Parker's model superseded, implicitly featured a non-vanishing pressure at infinity.\footnote{Indeed, the isothermal barometric law of \eqn{introbarometriclaw} has a non-vanishing pressure at infinity as the density in \eqn{introbarometriclaw} at $r=\infty$ is $\rho_o \exp(-\lp)$, and for barotropic flow, the pressure is a function only of density.  See Chamberlain (1963) for a detailed discussion of this breakdown of a barometric law using a kinetic theory approach.}  Parker reasoned that unless the pressure vanishes at infinity, a static atmosphere can only truly be held static if there is a finite inward pressure exerted on it at large radii.  In the case of the Sun, Parker pointed out that the vacuum-like conditions of the interstellar medium cannot possibly provide the necessary back pressure to keep the Sun's atmosphere in hydrostatic equilibrium (although see Velli 2001 and references therein for the extent to which this argument holds). 

More rigorously, we can exploit the TEP to identify a threshold temperature decline that determines whether or not an atmosphere can be held in hydrostatic equilibrium.  Integration of equation \eqn{staticequilb} over a non-isothermal atmosphere that extends from $r_o$ (where $\rho=\rho_o$ and $T=T_o$) to some radius $r$ can be written (Parker 1965),
\beq \rho(r)T(r) = \rho_oT_o\exp \bigb{-\int_{r_o}^r \f{\tau(r')}{r'}\,dr'} ,\label{rhoTeqn}\seq
where we have taken the pressure as $P(r) = c_s(r)^2 \rho(r)$.  In order for \eqn{rhoTeqn} to describe a static atmosphere surrounded by vacuum, the density must vanish at infinity, implying that the integral inside the exponent must be divergent at large $r$.  Conversely, a Parker wind is the steady equilibrium state of an atmosphere if the integral is convergent.  In terms of the TEP, $\rho$ vanishes at infinity if $\tau$ is an increasing function of $r$, while the density tends to a finite value if $\tau(r)$ is decreasing.  Physically, therefore, a spherical, \textit{isolated} static atmosphere is possible only if the magnitude of the gravitational potential energy of the gas outweighs its thermal energy at large radii.  For the critical case in which these energies are in balance, i.e. when $\tau(r)$ is constant, we see from \eqn{TEPdef} that the temperature profile satisfies $T(r)/T_o = r_o/r$ (provided $\lp$ does not vary with $r$) and from \eqn{rhoTeqn} that the integral diverges logarithmically.  

It seems to have been overlooked previously that there are no transonic Parker wind solutions with a $1/r$ temperature dependence, which occurs when $\gamma = 3/2$ (see \apx{newcritpts}).  
Importantly, $\gamma = 3/2$ is the critical adiabatic index that divides the behavior and solution space of transonic polytropic Parker winds.  For $\gamma<3/2$, transonic Parker winds are accelerating, while for $\gamma>3/2$, they are decelerating.  As summarized in \tbl{1}, the decelerating wind regime permits isolated hydrostatic solutions (which have $d\tau/dr > 0$), while only Parker winds can have a vanishing pressure at infinity for $1 \leq \gamma < 3/2$.  It is clear from \tbl{1} that the parameter space, $(\lp,\gamma)$, leading to spherically symmetric transonic Parker wind solutions is coupled in a simple way.  In \sec{discussion}, these HEP bounds are generalized to account for the effects of rotation. 

\begin{table*}
%\scriptsize
\begin{center}
\caption{Parameter Space of Spherically Symmetric Parker Wind Solutions.}
\begin{tabular}{c c c c c}  \\
\hline
\hline 
Polytropic & Permitted & TEP & Hydrostatic$^\dagger$ & Transonic \\
Index & HEP Range & Behavior & solutions? & solutions?\\
\hline
$\gamma=1$&$[2,\infty]$  & $d\tau/dr< 0$ &  No & Yes \\
$1<\gamma<3/2$&$[2,1/(\gamma-1)]$  & $d\tau/dr< 0$ &  No & Yes \\
$\gamma = 3/2$& $\lp = 2$	 & $d\tau/dr = 0$ & Yes & No   \\
$3/2<\gamma<5/3$& 	$[1/(\gamma-1),2]$ & $d\tau/dr > 0$ &	Yes & Yes   \\
$\gamma=5/3$& $[1.5,2]$ & $d\tau/dr > 0$ &	Yes & No   \\
\hline \hline 
\end{tabular}
 \end{center}
 $^\dagger$This refers to `isolated' hydrostatic solutions, i.e. those with a vanishing density at infinity.
\normalsize
\label{parspace}
\end{table*}

\subsection{The Applicability of Parker-like Disk Winds}
Magneto-centrifugally driven winds are often invoked as candidate mechanisms for explaining outflows from accretion discs.  In systems or regions of systems where magnetic forces might be dynamically unimportant, thermal driving is a likely contender (e.g., Proga 2007 \& references therein).  Just as Parker winds can be useful for modeling outflows from any spherical astrophysical body thought to be hot enough to exhibit a non-explosive, thermal expansion of gas, the Parker-like disc winds addressed in this paper can be used to model thermally driven winds from the coronae of accretion discs associated with AGN, X-ray binaries, unmagnetized protostellar discs, and protoplanetary discs.  Due to the diversity of physical scales in these systems, a preliminary step for constructing a 1D model is to identify a characteristic radius for invoking thermal driving, in order to calculate the HEP.  
First, it is worth emphasizing that the escape velocity for discs varies with distance $r_o$ along the disc midplane, so the HEP must be considered a function of $r_o$.  In this regard there is an intrinsic difference between 1D disc wind models and 1D spherical wind models.  
Namely, for a given mass $M_*$ and characteristic launching radius $r_o$ for a star or planet, varying the HEP samples different temperatures of the stellar corona or planetary exosphere.  Meanwhile, varying the HEP for a given central object mass for disc winds corresponds to altering either the temperature at a fixed distance along the midplane or the distance at a fixed temperature -- or both.

\subsubsection{The Thin-Disc Assumption}\label{thindisc}
The scale height of an isothermal corona is given by $H = \sqrt{2/\lp} r_o$, implying that our models can only be applied to regions of a \textit{flared} disc where $\lp >> 2$, as we are implicitly supposing that $H/r_o << 1$.  In order for our disc wind solutions to not be restricted to $\lp >> 2$, the sound speed within the disc must be considered separate from the sound speed at the base of the wind.  In other words, we imagine a cold, thin disc that acts as a reservoir of material capable of sustaining a wind.  Models of the internal disc structure show it to be complex and turbulent (e.g., Balbus \& Hawley 1998; Miller \& Stone 2000; Proga \& Begelman 2003; Turner et al. 2003; Hirose et al. 2006; Blaes et al. 2007; Krolik et al. 2007).  Since there is no analytic model for this dynamic internal structure of the disc, we cannot incorporate this complexity into our treatment.  Hence, the gas in the disc need not `match' onto the base of the streamlines of the heated surface gas that forms the outflow, as the situation is analogous to the photosphere-corona transition.   

\subsection{Sources of Heating for Thermal Driving}\label{heating}
The two simplest heating mechanisms believed to be capable of launching thermal winds in accretion discs are Compton heating and photoionization.  Here we discuss how to approximate the different wind regimes identified by BMS83 using the two input parameters $(\lp,\gamma)$ of Parker winds.

\subsubsection{Compton Heating}
The relevant length scale for AGN and X-ray binary disc winds is the Compton radius, the radius where the gravitational and thermal pressures are equal:
\beq R_{IC} \equiv \f{GM_*}{c_{IC}^2} = \f{GM_*m_p\mu}{kT_{IC}} .\seq
Here, $c_{IC}$ is the isothermal sound speed for gas heated to the inverse Compton temperature $T_{IC}$ (which can be $\sim 10^8\,K$ depending on the spectrum of radiation), defined by
\beq kT_{IC} = \f{1}{4}\avg{h\nu} ,\seq
where $\avg{h\nu}$ is the average photon energy from an isotropic radiation source of luminosity $\mathcal{L}$, namely
\beq \avg{h\nu} = \f{1}{\mathcal{L}} \int_0^\infty h\nu \mathcal{L}_\nu \,d\nu .\seq
As discussed by BMS83, regardless of magnitude of the luminosity, at radii beyond $R_{IC}$ the gas cannot remain quasi-static; the corona is itself unbound and better described as a vigorous wind region.  Woods et al. (1996) found the dividing cutoff for weak outflows to lie at the smaller radius $r_o \sim 0.1 R_{IC}$ (see also Proga \& Kallman 2002).  In terms of $\xi \equiv r_o/R_{IC}$, the HEP is 
\beq \lp = \f{1}{\gamma \xi}\bigp{\f{T_{IC}}{T_o(\xi)}}.\label{HEPbms83}\seq  

Depending on the luminosity, the wind regions to either side of $\sim 0.1 R_{IC}$ are further divided; BMS83 identified five solution regimes in all (see also Woods et al. 1996).  Each has an associated mass flux density, determined by $\xi$ and $\mathcal{L}/\mathcal{L}_{cr}$, where $\mathcal{L}_{cr}$ is a critical luminosity defined by
\beq \mathcal{L}_{cr} = \f{1}{8 \mu} \bigp{\f{c}{c_{IC}}} \bigp{\f{m_e}{m_p}} \mathcal{L}_E .\seq
Here, $\mathcal{L}_E$ is the Eddington luminosity, $c$ the speed of light, and $m_e$ the electron mass;  $\mathcal{L}_{cr}/\mathcal{L}_E < 0.1$ for $T_{IC} \ga 10^7$ K, so that thermal pressure dominates radiation pressure.  Parker-like disc winds are applicable in the regions affected by gravity, which includes the two weak wind regions with $\xi < 1$, labelled D and E by BMS83 with $\mathcal{L}/\mathcal{L}_{cr} < 1$ and $\mathcal{L}/\mathcal{L}_{cr} > 1$, respectively, as well as the portion of the `gravity inhibited' strong wind region, labelled C, with $\xi > 1$ and $\mathcal{L}/\mathcal{L}_{cr} << 1$.  The remaining two regions A \& B have high enough luminosities that gravity is dynamically unimportant and adiabatic losses insignificant in the subsonic flow regime. 

Utilizing Parker-like disc winds in the context of Compton heating amounts to a significant simplification of the theory developed by BMS83.  However, our models may provide an adequate approximation of the disc wind dynamics because the functional form for how the mass flux density scales with radius $r_o$ along the disc plane is identical with that found by BMS83 for both isothermal $(\gamma = 1)$ and isentropic $(\gamma = 5/3)$ flow.  We can account for this agreement by contrasting the two approaches used to treat the thermodynamics. 

The simplicity of invoking Parker winds resides in the use of a polytropic EoS ($P\propto \rho^\gamma$, where $P$ is the pressure and $\rho$ is the density), the conventional means for bypassing the heat equation when the source of heating is very complicated or poorly understood (see, e.g., Tsinganos \& Trussoni 1990; Sauty et al. 1999; Meliani et al. 2004) -- not the case with Compton-heated Corona, in which the thermodynamics can be conveniently handled via an entropy equation (BMS83).  If it is assumed that no heat is transferred via conduction or viscous dissipation to or from outflowing gas, conservation of energy dictates that the entropy production is proportional to the heating rate, $\Gamma$.  For optically thin gas heated to temperatures $T \ga10^6\,K$, the net heating and cooling rate is proportional to the difference $\Delta T = T-T_{IC}$ (e.g., Krolik et al. 1981).
We therefore see that when the heating rate is high throughout the entire subsonic wind region, so that there is a near balance of heating and cooling ($\Delta T = 0$), then an isothermal Parker wind with $T = T_{IC}$ will be a good approximation to BMS83's strong gravity, nearly isothermal region E (see \sec{polymdot} for details). 

In the opposite case of adiabatic, isentropic flow ($\gamma=5/3$), the entropy production is zero.  In the framework of BMS83, this can effectively occur when the heat-transport can altogether be ignored ($\Gamma \approx 0$), meaning that the heating time-scales are long compared to the flow time-scales.  More specifically, $\gamma=5/3$ applies to gas with no internal degrees of freedom that is heated to a high temperature $T \la T_{IC}$ in, say, a thin layer above the optically thick disc, that from there expands outward, loses additional pressure support upon being slowed by gravity, and thereby adiabatically cools.  In \sec{polymdot}, we explicitly show that the functional form of the mass flux density for a $\gamma = 5/3$ Parker wind is identical to the prediction given by BMS83 for their solution regime C.

\subsubsection{Photoionization Heating}
As discussed by BMS83, very similar physics underlies Compton and photoionization heating, albeit the cooling mechanism for the latter is significantly more complicated (line-cooling and recombination vs. inverse Compton).  Due to these complications, recent analytical studies have invoked a combination of numerical simulations and spherical Parker wind solutions in order to estimate global mass loss rates from protoplanetary discs (e.g., Gorti \& Hollenbach 2009, Owen et al. 2012). With the caveat mentioned in \sec{thindisc}, our models make it possible to move beyond a spherical wind boundary and analytically investigate Parker-like winds from the surface of the disc.  The starting point for our models is to make an explicit comparison between \eqn{HEPbms83} and the HEP for protoplanetary discs, 
\beq \lp = r_g/r_o, \seq
where $r_g = GM_*/c_o^2$ is `the gravitational radius', the distance where the gas becomes unbound because the escape velocity from the disc is equal to the thermal velocity of the gas.
For a constant temperature on the disc midplane (appropriate for a disc surface heated by EUV radiation to $\sim 10^4\,K$ or Compton heated to $T_o(\xi) \sim 10^7$ K), $r_g$ is constant, and we see that photoevaporative winds are qualitatively similar to Compton heated winds in the sense that $r_g$ plays the role of $R_{IC}$.  In either case, $\lp$ decreases as $r_o^{-1}$ due to the reduced escape speed. 

\subsection{Disc Streamline Geometry} \label{models}

\begin{figure*}
	\centering
	\includegraphics[scale=0.45]{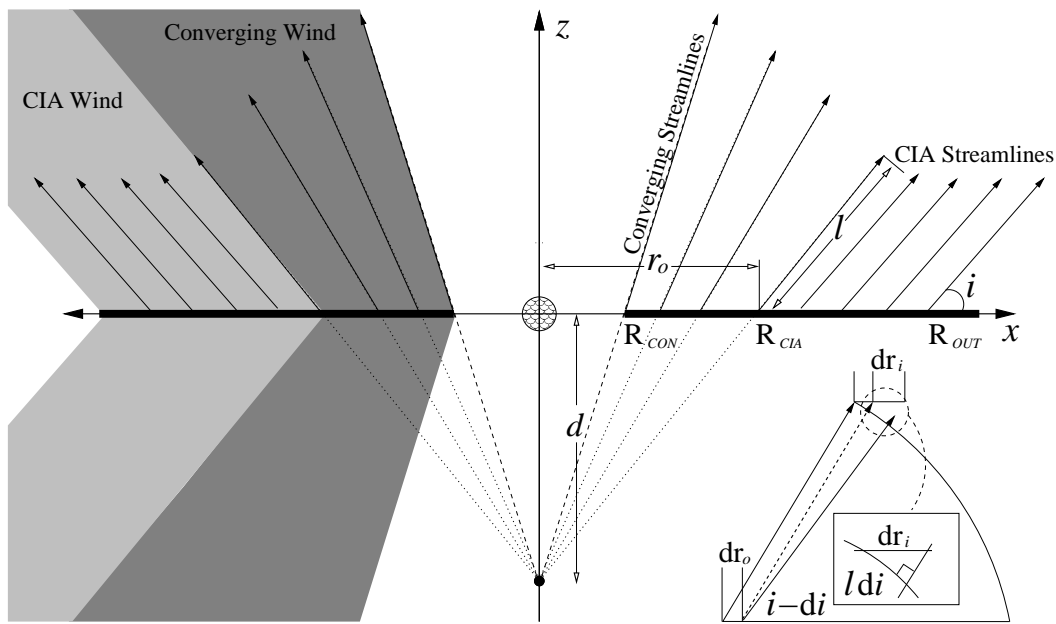}
	\caption{The biconical outflow of our `global' disc wind model is shown schematically on the left.  On the right, we depict our coordinate system and the streamline geometry of the CIA and Converging models.  In the fourth quadrant, we illustrate the geometry used to arrive at an expression for $A(l)$, which can be visualized as the area swept out around the $z$-axis by any two neighboring streamlines at a fixed distance $l$.  This notion becomes exact in the differential limit of closely spaced streamlines.}
	\label{modelgeometry2}
\end{figure*}

There are two routes to take in regards to specifying a geometry when finding solutions to wind equations (e.g., BMS83; Tsinganos \& Sauty 1992): (i) assign some trajectory to the flow emanating from the disc or (ii) self-consistently solve for that trajectory.  The first entails that an expression be provided for the flow tube area $A$ that enters the steady state continuity equation $\rho v A = \,constant$.  By adopting the geometry of the models discussed in the introduction, we necessarily take route (i).  As shown in  \fig{modelgeometry2}, our geometry is comprised of streamlines that are straight lines in the $(x,z)$-plane.  By rotating these straight streamlines around the $z$-axis, the actual trajectory traversed by the gas as it rises above the disc can be visualized; it spirals about a cone that widens according to the inclination angle $i$.  Besides being observationally motivated, a biconical flow area is the simplest possible choice, for the distance along a streamline $l=l(x,z)$ can be used as the sole variable instead of seeking some relationship between the cylindrical coordinates $x$ and $z$.  Our coordinates are related by $x=r_o + l \cos i$, $z=l \sin i$, and $r = \sqrt{r_o^2 + l^2 + 2lr_o\cos i}$.  CIA streamlines are self-similar, while Converging streamlines are not.  We find $A(l)$ for each configuration in \sec{ctyciacon}.

BMS83 and Fukue (1989) also took route (i) by assuming a flow configuration.  Fukue (1989) adopted an area function similar to that used by BMS83, but he did not self-consistently implement the polytropic EoS when he followed BMS83 in requiring that the wind be launched from rest from the disc midplane.  (No restriction is imposed on $v_o$ for the entropy equation used by BMS83.)  In our notation, BMS83 chose a generic area function aimed at parametrizing the streamline divergence: $A(l)=(1+l/r_o)^q$.  The parameter $q$ is constrained to lie between 0 (vertical flow) and 2 (spherical flow).  In \sec{ctyciacon}, we show that the Parker and Converging models have $q=2$, while the CIA model has $q=1$.   
 
Alternatively, route (ii) can be followed, in which an attempt can be made to calculate $A(l)$ as part of the solution.  This involves either solving the fully 2D, 2.5D, or 3D problem using numerical techniques or it requires making an extra assumption, such as self-similarity or force balancing.  The latter route was taken by Takahara et al. (1989), who arrived at an expression for $A(l)$ by assuming that the centrifugal force balances the component of the gravitational force perpendicular to the flow velocity at every distance $l$ along the streamline.\footnote{In the MHD literature, balancing forces perpendicular to the streamlines leads to the Grad-Shafranov (or transfield) equation.}  As discussed by BMS83, this is a valid approximation regardless of the streamline trajectory \emph{close do the disc midplane} if angular momentum is conserved.  It was soon pointed out by Fukue \& Okada (1990) that Takahara et al. (1989) misrepresented the gravitational force in calculating this force balance.\footnote{Takahara et al. (1989) used the x-component of the gravitational force rather than the component perpendicular to the streamline, to arrive at a self-similar streamline trajectory given by $z = x\sqrt{(x/r_o)^{2/3}-1}$.}
The correct expression for $A(x,z)$, obtained by Fukue \& Okada (1990), is significantly more complicated; it suffices to consider the shape of their streamlines in the $(x,z)$-plane.  At every footprint distance $r_o$, the local streamline is found from
\beq r = \f{r_o + cz}{2}\bigb{1 +\sqrt{1- \bigp{\f{2z}{r_o + cz}}^2} } ,\label{Fukuestreamlines}\seq
where $r=\sqrt{x^2 + z^2}$ is the spherical position coordinate.  The parameter $c$ is an integration constant that resulted from the correct treatment of the force balance, which entailed solving an ordinary differential equation for $dx/dz$.  Hence, $c$ is related to the slope -- the opening angle of the streamline -- and defines a family of streamlines at every footprint radius $r_o$.  See Figure 1 in Fukue \& Okada (1990) for a plot comparing their streamlines for various values of $c$ with the self-similar streamlines found by Takahara et al. (1989).  By observing that the concavity or convexity of the streamlines in \fig{stefans} is pronounced only near the midplane, we can conclude that using \eqn{Fukuestreamlines} would not be an improvement over our choice of geometry, as \eqn{Fukuestreamlines} does not capture this feature.

Nevertheless, the findings of Fukue \& Okada (1990) provide a physical argument supporting our choice of geometry.  Examination of \eqn{Fukuestreamlines} reveals that close to the disc midplane, their streamline function is indeed just a straight line.\footnote{To see this, note that to first order in $z$, when $z<<r_o$, $r\approx x$ and the right hand side of \eqn{Fukuestreamlines} is $\approx r_o + cz$.  Hence, $x = r_o + cz$, which is just our $x$-coordinate, provided we identify the constant as $c=1/\tan i$.  Thus, $c\geq2$ implies $i \leq \tan^{-1}(.5)=26.6^\circ.$}\label{FO}  In other words, balancing the gravitational and centrifugal forces perpendicular to the flow \emph{implies} straight streamlines close to the $x$-axis.  Equivalently, conical streamlines define the path of minimum effective potential near the disc midplane.  Moreover, Fukue \& Okada showed that streamlines curve back on themselves (and intersect the $z$-axis) if $c < 2$.  For streamlines to extend to infinity, it is required that $c\geq 2$.  This is the requirement that the inclination angle $i \la27^\circ$ (see footnote 6), which is approximately the opening angle of the self-similar streamlines obtained by Luketic et al. (2010) and shown in \fig{stefans}.  Note, however, that the square root in \eqn{Fukuestreamlines} spoils the self-similarity that the streamlines would possess if the radical were zero.

One would expect that a model featuring streamline curvature in the $(x,z)$-plane would lead to significantly different wind solutions if the area function $A(l)$ directly determined the critical point location.  This is not the case, however, as the well known rocket-nozzle analogy in stellar wind theory revealed that (the effective) gravity, more so than the flow tube area, plays the role of the converging-diverging nozzle to facilitate a transition from subsonic to supersonic flow (e.g., Lamers \& Cassinelli 1999; we define the equivalent nozzle function in \sec{equivnozzle}).  To stress this point, consider how the area term enters the equation of motion per elimination of the density gradient (by taking a logarithmic derivative of the continuity equation),
\beq \f{1}{\rho}\f{d\rho}{dl} = -\f{1}{v}\f{dv}{dl} -\f{1}{A}\f{dA}{dl} \label{rhogradient}. \seq
The first term on the right hand side exemplifies the outcome of adopting a fluid treatment: the density gradient (and hence the pressure gradient for a polytropic EoS) itself depends on the flow acceleration.  It is this term that gives rise to a singularity upon solving the equation of motion for $dv/dl$.  In turn, the second term, $d\ln A/dl$, which is more a measure of streamline divergence than of the area between streamlines, influences the position of this singularity -- the location of the critical point.  This location would not change by much had we analytically modeled the exact area in \fig{stefans}, as azimuthal streamline divergence is well accounted for using straight streamlines. 

%===============================================================================
% FORMULATION
%===============================================================================

\section{Hydrodynamic formulation} \label{formulation} 
In this section we present a general formalism for solving 1D thermally driven wind equations under either spherical or axial symmetry.
We adopt all of the simplifications of the classic Bondi and Parker problems, namely we consider the hydrodynamic limit, assuming a single-fluid treatment and inviscid, barotropic flow.  Imposing these restrictions allows the vector Eulerian momentum equation to be integrated and the problem solved using a simple Bernoulli function, constant along a given streamline.  The forces acting on a fluid element are: the force of gravity from a central source, gas pressure, and the centrifugal force when there is nonzero rotation.  We use the conventional polytropic EoS in lieu of the energy equation.  We only relax the assumption of spherical symmetry by adding an azimuthal velocity component, conserving the specific angular momentum of the fluid, and we allow for arbitrary amounts of streamline divergence.  

Our formulation is an extension of the classic isothermal and polytropic Parker problems into cylindrical symmetry.  Because of the equivalence of wind and accretion equations, our problem is also a generalization of the classic Bondi problem (Bondi 1952).  Bondi's analysis entailed applying boundary conditions at infinity, where both the velocity and gravitational potential vanish.  Our solution allows boundary conditions to be applied at any finite distance away from the central gravitating object.  Although we do not address the generalized Bondi problem because our focus is on winds, it should be kept in mind that any explicit reference to boundary conditions taken at `the base' -- be it the disc midplane or the surface of the central object -- can equally well denote `outer' boundary conditions appropriate to accretion problems. 

\subsection{The Continuity Equation}\label{ctyciacon}
In \apx{ctyeqn}, we show that for the geometry of \fig{modelgeometry2}, the steady-state continuity equation can be written as
\beq d\dot{M}=\rho(l)v(l)A(l) ,\label{dmdot}\seq
where $d\dot{M}$ is the differential mass-loss rate at the location $r_o$.  The area between streamlines, $A(l)$, can be determined from \eqn{streamlineA}, after specifying $di/dr_o$, the adjacent streamline divergence.  
The CIA model has no streamline divergence ($di/dr_o = 0$), giving
\beq A(l)= 2\pi dr_o(r_o + l\cos i)\sin i. \label{ciaarea}\seq
We see by \fig{modelgeometry2} that the total midplane area occupied by the CIA model, obtained by letting $l=0$, $\sin i = 1$, and integrating over $dr_o$ from $R_{\,CIA}$ to $R_{\,OUT}$ is correctly given as $\pi(R_{\,OUT}^2-R_{\,CIA}^2)$.   

The Converging model, with geometry obeying $\tan i = d/r_o$, has streamline divergence $di/dr_o = -\cos i \sin i/r_o$, so
\beq A(l)= \f{2\pi dr_o(r_o + l\cos i)^2\sin i}{r_o}. \label{conarea}\seq
Notice that for the same $r_o$, both the CIA and Converging models have a differential base area given by $A_o \equiv A(l=0) = 2\pi r_o dr_o \sin i$. 

\subsubsection{The Parker Model from the Converging Model with $i=0^\circ$}
Only ratios of the area appear in the equations governing the flow, as in 
\beq \f{1}{A(l)}\f{dA(l)}{dl}= \bigp{\f{q\cos i}{r_o + l\cos i}} ,\seq  
where $q=1$ for the CIA model and $q=2$ for the Converging model.  The flow quenching factor $\sin i$ does not enter the disc wind problem except when calculating $\dot{M}$.  We see, therefore, that 
the Converging model contains the Parker model as a special case, for when $i=0^\circ$, $d=0$, bringing the converging point to the source of gravity.  Then $r_o + l$ is just the spherical coordinate $r$, $r_o$ representing the coronal radius rather than the footprint distance.  The only distinction that needs to be made is that equation (\ref{conarea}) formally does not apply in that case since $A(l)=0$ -- in cylindrical symmetry, there is no width between streamlines because they all overlap on the x-axis.  This can be thought of as a collapse to spherical symmetry, so the Parker model results, albeit with the adjustment that the differential base area becomes $A_o = 2\pi r_o^2\sin \theta d\theta$, where $\theta$ is the spherical polar coordinate, instead of $A_o = 2\pi r_o dr_o \sin i$.  

\subsubsection{Cylindrical Parker Winds: The CIA model with $i=0^\circ$}\label{CylindricalParker}
Since the Parker wind solution is recovered from the Converging model at $i=0^\circ$ (and with zero rotation), it is reasonable to ask if the solution to the CIA model at $i=0^\circ$ bears any significance.  It turns out that this solution was obtained by Skinner \& Ostriker (2010) and included as a testbed problem in their extension of the MHD code \emph{Athena} into cylindrical coordinates.  This `cylindrical version' of a rotating Parker wind, as they referred to it, can be viewed as a wind flowing perpendicular to the symmetry axis of evenly spaced concentric cylinders (with $A_o = 2\pi r_o dz$).  Skinner \& Ostriker's (2010) rotating wind test demonstrated cylindrical Athena's ability to maintain steady state, transonic flows and conserve angular momentum in cylindrical symmetry.  Our solutions for both the Converging and CIA models with $i>0^\circ$ open up the possibility of allowing this test to incorporate the $z$-dimension. 

As we demonstrate, the proper procedure for wind equations is to take reference quantities at the footprint of a given streamline.  The equations obtained for the CIA model at $i=0^\circ$ by Skinner \& Ostriker (2010) are seemingly the same as ours, yet the problem as they pose it is poorly formulated because they used reference quantities defined at infinity, where the pressure (and hence sound speed for a polytropic EoS) vanishes.  Specifically, they normalized the Bernoulli function to $c_\infty^2/(\gamma -1)$; their solutions do not suffer from this choice due to their assigning a value to the Bernoulli constant (thereby setting the location of the sonic point) a priori.  

\subsection{The Bernoulli Function}
By Bernoulli's theorem, the Bernoulli function is a constant on a streamline:
\beq \f{1}{2}v^2 + \Phi + h =constant, \label{Bernoulli}\seq for enthalpy $h=\int d\,P/\rho$ 
and bulk flow velocity $v = \sqrt{v_x^2 + v_\phi^2 + v_z^2} = \sqrt{v(l)^2 + v_\phi^2}.$  
We denote the Bernoulli constant as $B_o$ and emphasize that, while it is to be evaluated at the boundary, 
\beq B_o \equiv \bigp{\f{1}{2}v^2 + \Phi + h} \:\bigg\lvert_{BDY}\:, \label{Bo}\seq
it is a priori unknown because $v(l=0)$ is unknown.

To define $B_o$, both the temperature $T$ and density $\rho$ at the base of every streamline must be specified.  We take these quantities to be $T_o$ and $\rho_o$, respectively.  For an ideal gas EoS, this is equivalent to specifying the pressure at at every footprint location $r_o$.
The barotropic assumption, $dP = (\partial{P}/\partial\rho) d\rho$, is satisfied by an ideal gas EoS combined with the polytropic fluid relation, $T = T_o(\rho/\rho_o)^{\gamma-1}$.  Explicitly we have $P = \rho_o k T_o(\rho/\rho_o)^{\gamma}/\mu m_p$, which gives
\beq
 h = \int\f{1}{\p}\d{P}{\p}d\p= \f{c_o^2}{\p_0^{\gamma-1}}\int \rho^{\gamma-2}\,d\rho =\\
  \begin{cases}
    c_s^2 \ln(\rho/\rho_c)       & \text{if } \gamma =1 \\
   c_s^2 /(\gamma-1) & \text{if } \gamma > 1 ,
  \end{cases} \label{hcases}
 \seq 
provided that we absorb into $B_o$ the constant term $c_s^2 \ln(\rho_c)$ for $\gamma=1$, as well as the constants of integration.

For rotational motion in a plane under a central force, the specific angular momentum $L$ is a constant of the motion: $L= xv_\phi$, where $x=r_o+l\cos i$.  We will present our disc wind results for a disc rotating at Keplerian velocities, in which the disc angular velocity at any location $r_o$ is $\Omega_K=\sqrt{GM_*r_o}/r_o^2$.  However, for treating Parker winds we follow Keppens \& Goedbloed (1999) in allowing for arbitrary rotation rates $\Omega$, parametrizing $v_\phi$ on the equatorial plane by some factor $\zeta$ of the adiabatic sound speed at the base, i.e. $v_\phi(l=0)=\Omega r_o= \zeta c_o$, giving 
\beq v_\phi = \Omega r_o \bigp{\f{r_o}{x}}=\zeta c_o \bigp{\f{r_o}{r_o + l\cos i}}. \seq
Keplerian rotation corresponds to $\zeta = \sqrt{GM_*/r_o c_o^2}=\sqrt{\lp}$.
Rotation therefore enters the problem as an effective potential,
\begin{align}
\Phi \rightarrow \Phi_{eff} &= -\f{GM_*}{r} + \f{v_\phi^2}{2} \nonumber \\
& = -\f{GM_*}{\sqrt{r_o^2 + l^2 + 2lr_o\cos i}} + \f{\zeta^2 c_o^2}{2} \bigp{\f{r_o}{r_o + l\cos i}}^2 .
\end{align}
For $\gamma > 1$ then, the Bernoulli function reads
\beq B_o = \f{1}{2}v(l)^2 + \Phi_{eff}(l) + \f{c_s(l)^2}{\gmone}.\label{Beqn}\seq
We treat the isothermal ($\gamma=1$) case in \sec{LambertWfctsoln}.

\subsection{The Equation of Motion}
The equation of motion, which we will sometimes refer to as $F(l,v,dv/dl)=0$, contains the velocity gradient.  Critical points arise whenever the velocity gradient becomes undefined, i.e. when $dv/dl = 0/0$.  Hence, the Bernoulli function must be accompanied by $F(l,v,dv/dl)=0$ to seek out these critical points.  $F(l,v,dv/dl)=0$ is found by first differentiating \eqn{Beqn} to give
\beq \d{B_o}{l}= v\d{v}{l} + \d{\Phi_{eff}}{l} + \f{c_s^2}{\p}\d{\p}{l} = 0 ,\label{Fprelim}\seq
and then by eliminating the density gradient using the continuity equation.  The relevant derivative is given in equation (\ref{rhogradient}).  Further dividing by $c_s^2$ gives 
\beq F(l,v,dv/dl) \equiv\left(1-\frac{c_s^2}{v^2}\right)\f{v}{c_s^2}\d{v}{l} + \f{1}{c_s^2}\d{\Phi_{eff}}{l} - \frac{1}{A}\d{A}{l} =0.\label{F}\seq

\subsection{Dimensionless Formulation} \label{Dimensionless}
Our disc wind problem depends on a total of three parameters, namely, $\lp$, $\gamma$, and $i$.  We find it natural to normalize distances to the gravitational radius,
\beq r_g = \lp r_o = \f{GM_*}{c_o^2}. \seq
Justification for this choice is obtained by shifting one's viewpoint to consider all subscripts `$o$' as standing for `outer' rather than midplane boundary conditions.  Then in the limit $c_o\rightarrow c_\infty$, $r_g$ is the so-called Bondi length.  Since we will discuss results for both spherical winds and disc winds, we will differentiate disc wind bulk velocities by normalizing to $V_{esc} = \sqrt{GM_*/r_o}$, the escape velocity from a thin Keplerian disc at a distance $r_o$ along the disc, instead of $v_{esc} = \sqrt{2}V_{esc}$.  The HEP has several different guises in terms of these characteristic quantities, namely
\beq \lp  = \f{V_{esc}^2}{c_o^2} = \f{v_{esc}^2}{2c_o^2}  = \f{r_g}{r_o}.\label{discHEPdef}\seq 

\subsubsection{Unknown Critical Point Quantities}\label{lcparameter}
We introduce a quantity analogous to $\lp$, defined by
\beq\lc \equiv \f{V_{esc}^2}{c_s(l_c)^2},\label{lc} \seq
where $c_s(l_c)$ is the sound speed at the critical point (subscripts `c' will be used to denote quantities evaluated at the sonic point throughout).  Since $c_s(l_c)$ is in a one-to-one relationship with the critical point distance $l_c$, $\lc$ is a central unknown.  In general, $\lc$ can only be solved for numerically.  Many quantities of interest such as the mass loss rate, the initial velocity, and the terminal velocity can be simply expressed in terms of the ratio $\lc/\lp = T_o/T_c$.  Note that $\lc=\lp$ in the isothermal case.  For $1<\gamma \leq 5/3$, the ratio $\lc/\lp$ is equal (by construction) to the fundamental constants of the problem,
\beq \f{\lp}{\lc} = \f{(B_o/c_o^2)}{e_c} , \label{lc2} \seq
where $e_c = e_c (\x_c)$ is the critical point constant, and is formally given by
\beq e_c = \f{B_o}{c_s^2(l_c)}.\seq
We emphasize that $e_c$ is a constant determined independently of $B_o$ (see \S\ref{critptconst}).  
 
\subsubsection{Dimensionless Equations}
We now rewrite the governing equations into a form suitable for numerical implementation.   We begin by introducing the following dimensionless variables:
\begin{eqnarray*} 
\text{distance along a streamline: \hspace{4 pt}} \x &=& \f{l}{r_g}  \label{xi} ,\\
\text{specific kinetic energy: \hspace{4 pt}} y &=& \f{1}{2}\f{v^2}{c_s(\x_c)^2}  \label{K} ,\\
\text{Mach number: \hspace{4 pt}} \mathcal{M} &=&\f{v}{c_s}  \label{M} ,\\
\text{Mach number squared: \hspace{4 pt}} w &=& \mathcal{M}^2  \label{w} ,\\
\text{sound speed squared: \hspace{4 pt}} s &=& \f{c_s^2}{c_s(\x_c)^2}  \label{s} .
\end{eqnarray*}
Here, $\x_c = l_c/r_g$ is the dimensionless critical point distance.  The variables $y$, $w$, and $s$ are related by
\beq y = \f{sw}{2} .\label{y}\seq
We prefer simply keep $\rho/\rho_o$ and $A/A_o$ instead of renaming the density and flow tube area.  
Then the continuity equation becomes
\beq \f{\dot{m}}{\p_o c_s(\x_c)} =  \sqrt{2y}\f{\mathrm{A}}{\mathrm{A}_o}\f{\rho}{\rho_o}, \label{CeqnD}\seq
where the mass flux density, $\dot{m}$, is defined as
\beq \dot{m}\equiv\f{d\dot{M}}{A_o}.\seq
Note that $\dot{m} = \rho_o v_o = \p_o c_o \mathcal{M}_o$, where $\mathcal{M}_o$ is the initial Mach number.
The polytropic relation is now
\beq s = s_o\bigp{\f{\rho}{\rho_o}}^{\gmone} =\f{\lc}{\lp}\bigp{\f{\rho}{\rho_o}}^{\gmone}, \label{seqnD}\seq
where by construction $s_o \equiv s(\x=0) = \lc/\lp$.  

Dividing \eqn{Beqn} by the unknown quantity $c_s(\x_c)^2$ gives the dimensionless Bernoulli function,
\beq  e_c = y + \f{\lc}{\lp}U_{eff} + \gfac s, \label{BeqnD2}\seq
where 
\begin{align}
U_{eff} &= U + U_{centrif} \nonumber \\
&= -\f{1}{\sqrt{\x^2 + 2\x f \cos i +  f^2}} + \f{1}{2}\bigp{\f{\zeta f}{f + \x\cos i}}^2, 
\end{align}
and we have introduced the following quantities:
\begin{eqnarray*} 
\text{gravitational potential: \hspace{4 pt}} U &=& \f{\Phi}{c_o^2},\\
\text{centrifugal potential: \hspace{4 pt}} U_{centrif} &=& \f{v_\phi^2}{c_o^2}  ,\\
\text{inverse HEP: \hspace{4 pt}} f &=& \f{1}{\lp} .
\end{eqnarray*}
The above equations take their simplest form by eliminating any reference to $\rho_o$ and $A_o$.
Expressed this way, the equations depend only on the ratio $\lc/\lp$ and values taken at the critical point, making it clear that the accretion equations (with outer boundary condition $\lp \rightarrow 0$ at $A_o \rightarrow \infty$ but with finite $\lc/\lp$ and $A_c$) are identical to the wind equations.  First we square \eqn{CeqnD} and express it in terms of $w$: $\dot{m}^2 = s w (\p_o c_o)^2(\lc/\lp)(\mathrm{A}/\mathrm{A}_o)^2 (\rho/\rho_o)^2$.  The polytropic EoS, \eqn{seqnD}, permits substitution for $\rho/\rho_o$.  Further evaluating $\dot{m}^2$ at the critical point where $s_c=w_c=1$ yields the combined continuity equation/polytropic EoS in terms of $\dot{m}_c^2$,
\beq \dot{m}^2=\dot{m}_c^2\f{\mathrm{A}^2}{\mathrm{A}_c^2}ws^{\gfacb} .\label{CPeqn} \seq
Defining 
\beq \Lambda \equiv \f{\dot{m}}{\dot{m}_c} ,\seq
\eqn{CPeqn} becomes
\beq \Lambda^2 = \f{\mathrm{A}^2}{\mathrm{A}_c^2}ws^{\gfacb}. \label{eigval}\seq
Rewriting the dimensionless Bernoulli function in terms of $s$ and $w$ gives
\beq  e_c = \f{sw}{2} + \f{\lc}{\lp}U_{eff} + \gfac s, \label{BeqnD3}\seq
and equations \eqref{eigval} and \eqref{BeqnD3} together comprise an algebraic system of two equations for the two unknowns $s$ and $w$.  Once $e_c$ is evaluated, an explicit solution for $w$ can be found. 

The equation of motion, \eqn{F}, becomes
\beq F\equiv\left(1-\frac{s}{2y}\right)\f{1}{s}\d{y}{\x} + \f{1}{s}\f{\lc}{\lp}\d{U_{eff}}{\x} - \frac{1}{\mathrm{A}}\d{\mathrm{A}}{\x} =0. \seq
This can be further simplified by letting $y'=dy/d\x$, $\mathrm{A}' =  d\mathrm{A}/d\x$, and by defining the (minus of the) effective gravitational force as
\beq g = \f{dU_{eff}}{d\x} = \f{\x + f \cos i}{(\x^2 + 2\x f \cos i +  f^2)^\f{3}{2}} -\f{(\zeta f)^2\cos i}{(f + \x\cos i)^3} .\label{geff}\seq
$F(\x,y,y')=0$ now reads
\beq F\equiv (1-\frac{s}{2y})\f{y'}{s} + \f{g}{s}\f{\lc}{\lp} - \f{\mathrm{A}'}{\mathrm{A}} = 0 \label{F0}.\seq

\subsection{Critical Point Conditions}
\subsubsection{The Critical Point Constant} \label{critptconst}
The value of the critical point constant $e_c$ is found from \eqn{BeqnD3} evaluated at the critical point,
\beq e_c \equiv \f{sw}{2} + \f{\lc}{\lp}U_{eff} + \gfac s \Bigg\lvert_{\x=\x_c}.\label{egamma}\seq
Again since $s_c=w_c=1$, we have that
\beq e_c = \f{1}{2}\bigp{\f{\gpone}{\gmone}} + \f{\lc}{\lp}U_{eff}(\x_c) .\label{ecpoly}\seq
In general, therefore, $e_c$ depends on the critical point distance $\x_c$.  It is easily seen that despite the fact that the classic Bondi and Parker problems can have very different Bernoulli constants $B_o$, they both have the same value of $e_c=B_o/c_s(\x_c)^2$.  The singular nature of their equations at the critical point are identical.  In that spherically symmetric case, $e_c$ is independent of $\x_c$ and $\lc$:
\beq e_c = \f{1}{2}\bigp{\f{5-3\gamma}{\gmone}} \text{     (for spherical symmetry only)}  .\label{egbondi}\seq 
Rotation breaks this equivalence because then $e_c = e_c(\x_c)$, and $\x_c$ depends on the boundary conditions. 

\subsubsection{The Singularity and Regularity Conditions} \label{singnreg}
The singularity condition identifies all points at which the flow acceleration is undefined, i.e. all values of $\x$ for which $F(\x,y,y')=0$ is independent of $y'$: 
\beq \pd{F}{y'}=0 .\label{singularity}\seq
From \eqn{F0}, the set of possible points picked out by \eqn{singularity} are those that satisfy $y=s/2$ (or in physical units, $v=c_s$) at $\x_c$.
The regularity condition, in turn, defines the acceleration at this point as the slope of $F$ in the $(\x,y)$-plane: $y'=-(\partial F/\partial \x)/(\partial F/\partial y)$, or as it is more commonly stated,
\beq \pd{F}{\x} + y'\pd{F}{y} =0.\label{regularity}\seq
Equation \ref{regularity} is formally derived by ensuring that $dF(\x,y,y')/d\x=0$ all along the solution curve, which is equivalent to requiring a finite jerk, i.e. that $y''$ is bounded at $\x_c$ (Lamers \& Cassinelli 1999, \S8.7).  The role of the regularity condition is to ensure the continuity of the solution at the critical point.  Since this point coincides with the sonic point for our problem, it marks the region where the flow loses communication with what is happening downstream. In the neighborhood of this point then, there could potentially be thermodynamically different situations, which would result in a shock\ -- a discontinuity in $y'$.  Physically, therefore, the regularity condition prevents shocks, i.e. it demands that nothing special happens with the flow at the critical point.  For more complicated equations of motion, e.g. with line-driving included, explicit use of the regularity condition is required to determine the location of the critical point (e.g., CAK, Tsinganos et al. 1996).  For thermally driven winds, it is not needed, but we will make use of it in \sec{rotation} to interpret the negative root of the isothermal critical point equation.
 
\subsubsection{The Relation Between $\lc$ and $\x_c$}
The singularity condition combined with the equation of motion yields a relationship between $\x_c$ and $\lc$; it does not directly determine the location of $\x_c$ (except for the isothermal case when $\lc=\lp$).   With $y=s/2$ at the critical point, \eqn{F0} gives,
\beq \f{\lp}{\lc} = g_c\f{\mathrm{A}_c}{\mathrm{A}'_c}. \label{criteqn}\seq
Here, $\mathrm{A}_c/\mathrm{A}'_c = ( f  + \x_c \cos i)^q/(q \cos i)$, where $q=1$ for the CIA model and $q=2$ for the Parker ($i=0^\circ$) and Converging models.

\subsubsection{The Location of the Critical Point(s)} \label{critpteqn}
The innocuous looking equation $e_c = (\lc/\lp)B_o/c_o^2$ combined with \eqn{criteqn} determines the location of the critical point.  Equivalently, we can evaluate \eqn{BeqnD3} at the lower boundary,
\beq e_c = \f{\lc}{\lp}\bigb{\f{w_o}{2} + U_{eff,o} + \gfac} ,\label{ecbdy}\seq
where, using the definitions of $y$ and $s$, we have factored out $s_o=\lc/\lp$.  
A relation between $w_o$ and $\lc$ follows from \eqn{eigval}:
\beq w_o = \bigp{\Lambda \f{\mathrm{A}_c}{\mathrm{A}_o}}^2\bigp{\f{\lp}{\lc}}^{\f{\gpone}{\gmone}} \label{wo}\seq
With equations \eqref{wo} and \eqref{ecpoly} both substituted into \eqn{ecbdy}, and noting that $U_{eff,o} = -\lp+\zeta^2/2$, the general equation that must be satisfied by a critical point is
\begin{align}
&\f{\lp}{\lc}\bigb{\f{1}{2}\bigp{\f{\gpone}{\gmone}} + \f{\lc}{\lp}U_{eff}(\x_c)} =\\
&\bigb{\f{1}{2}\bigp{\Lambda \f{\mathrm{A}_c}{\mathrm{A}_o}}^2\bigp{\f{\lp}{\lc}}^{\f{\gpone}{\gmone}} -\lp + \f{\zeta^2}{2} + \gfac}. \label{critpointeqn}
\end{align}
All appearances of $\lp/\lc$ in \eqn{critpointeqn} are to be eliminated using \eqn{criteqn}.  The resulting equation can only be solved numerically -- with a root finder capable of detecting multiple roots -- except for the classic Bondi problem.  We solved \eqn{critpointeqn} for all of its roots using simple bracketing and bisection (Press et al. 1992) with a tolerance of $10^{-13}$ and a bracket spacing of $\Delta\x = 0.005$ (although a spacing of $10^{-3}$ was required for $\gamma \leq 1.1$).  We observed that \eqn{critpointeqn} almost always possess two roots.  For the classic Parker problem, the transonic\emph{ inflow} solution of the second root satisfies \eqn{wo} -- there are never two outflow solutions for the same set of parameters $(\lp,\gamma)$.  This issue is taken up again in \sec{topology} (and see \apx{newcritpts}).

\subsection{The Polytropic Fluid Solution} \label{solns}
Rearranging \eqn{BeqnD3} to isolate $s$ gives 
\beq \bigp{\f{w^{-1}}{\gmone}+\f{1}{2}}sw=e_c -  \f{\lc}{\lp}U_{eff} .\label{Bfors}\seq
An explicit solution to the problem, i.e. a solution comprised of separated functions of the dependent and independent variables, follows from substituting \eqn{eigval} solved for $s$ into \eqn{Bfors} and multiplying both sides by $\bigp{\Lambda\, \mathrm{A}_c/\mathrm{A}}^{-2\igfacb}$:
\beq F(w)  = \Lambda^{-2\igfacb}X(\x) , \label{explicit}  \seq 
where
\begin{align} 
F  & = \bigp{\f{w^{-1}}{\gmone}+\f{1}{2}}w^{\frac{2}{\gamma +1}}, \nonumber \\
X   & = \bigp{\f{\mathrm{A}}{\mathrm{A}_c}}^{2\igfacb}\bigp{e_c - \f{\lc}{\lp}U_{eff}} \nonumber .
 \end{align}
Note that an explicit solution cannot be found in terms of the kinetic energy ($y=sw/2$), as ridding the left hand side of \eqn{Bfors} of $\x$-dependence to give $F=F(y)$ is not possible.  Also note that insofar as there are solutions for which the ratios $\lc/\lp$ and $\mathrm{A}/\mathrm{A}_c$ are the same regardless of whether inner our outer boundary conditions will be applied, the explicit solution is completely independent of the boundary conditions.  Hence, inner or outer boundary conditions need to be applied separately to pick out the inflow or outflow solutions, respectively. 

\subsubsection{The Transonic Solutions}
Bondi (1952) explained why the maximum value of $\Lambda = \dot{m}/\dot{m_c}$ occurs for the branches that pass through the critical point.  We reiterate his logic linking the maximum mass flux density to the transonic solutions, since it follows from simple mathematical considerations.  For $\gamma >1$, both $F$ and $X$ are well-behaved functions with minimum values, so a solution ceases to exist when the right hand side of \eqn{explicit} becomes smaller than the minimum of the left hand side, $F_{min}$.  The product $\Lambda^{-2\igfacb}X(\x)$ is made smallest for some $X_{min}$ and $\Lambda_{max}$.  Therefore,
\beq \Lambda_{max} = \bigp{\f{F_{min}}{X_{min}}}^{\f{\gamma+1}{2(\gmone)}} .\seq
It is easily seen that $F_{min}$ occurs for $w=1$, i.e. $w = w_c$, at which value $F_{min} = .5(\gpone)/(\gmone)$.  Evidently, $X_{min} =F_{min}$ is found at $\x_c$, which can readily be verified because \eqn{criteqn} must hold at the critical point.  Thus,
\beq \Lambda_{max} \equiv \Lambda_c = 1. \seq
Solutions that occur for $\Lambda>1$ must accordingly have $X >X_{min}$, or equivalently $\x<\x_c$ or $\x>\x_c$ for all $\x$, to satisfy \eqn{explicit}.  These are the double-valued solutions discussed below.      

In practical terms, we have just shown that to obtain the transonic solutions, simply set $\Lambda=1$.  Of course, hindsight into the nature of the problem led to this convenient choice of $\Lambda$, also made by Holzer \& Axford (1970). 

Aside from the solution topology, our formulation of the general polytropic problem is now complete.  In \apx{formulae}, we provide formulae to compute all other variables and quantities of interest from the ones already given.  We apply our formalism to recover the solution of the Bondi problem in \apx{Bondiproblem}. 

\subsubsection{The Full Solution Topology: 4 Transonic Solutions!} \label{topology}
As Bondi also pointed out in his 1952 paper, $\Lambda$ acts as an eigenvalue in that each value of $\Lambda$ corresponds to a unique set of branches, or sets of points in the $(\x,w)$-plane that correspond to the two roots of the non-linear function $R(w,\x) \equiv F(w) - \Lambda^{-2\igfacb}X(\x) = 0$.  He was the first to show that the solution possesses an X-type topology containing both single-valued branches for $\Lambda < 1$ and double-valued branches for $\Lambda > 1$.  For $\Lambda<1$, one root defines a subsonic branch with $w < 1$ and the other a supersonic branch in which $w > 1$ everywhere.  This property makes the root-finding procedure for $w$ straight forward, since for a given $\x$, the two roots $w_1$ and $w_2$ are always bracketed by $0 < w_1 <1$ and $1<w_2<10+$.  As $\Lambda \rightarrow 1$ from slightly below $1$, these branches approach each other, bending ever more toward the critical point, and finally join each other at the single point $\x_c$ for $\Lambda =1$, thus forming the transonic solutions.  

For the double-valued $\Lambda>1$ solutions, one root defines a sub-critical branch with $\x < \x_c$ and the other a super-critical branch with $\x > \x_c$ always.  While these branches cannot represent viable wind or accretion solutions by themselves, they are crucial to forming complete physical solutions.  That is, these branches serve the purpose of allowing for shock transitions to inflows or outflows (Holzer \& Axford 1970; Theuns \& David 1991 and references therein).  For instance, a stable wind solution consists of a transonic $\Lambda=1$ velocity profile that matches onto a subsonic, super-critical branch upstream of a termination shock (Velli 1994, 2001). 

It has not been emphasized in the literature that the solution topology of the polytropic Parker problem differs from the Bondi problem in one important respect: there are \emph{two sets} of critical point solutions to choose from for every value of the adiabatic index in the range $1<\gamma<5/3$ (see \apx{newcritpts}).  As we already mentioned, \eqn{critpointeqn} almost always possesses two roots and each one yields a family of solutions with the topology described above.  Ultimately, therefore, there are two transonic outflow solutions to choose from for a given set of parameters.  

\begin{figure}
	%\centering
	\includegraphics[scale=0.55]{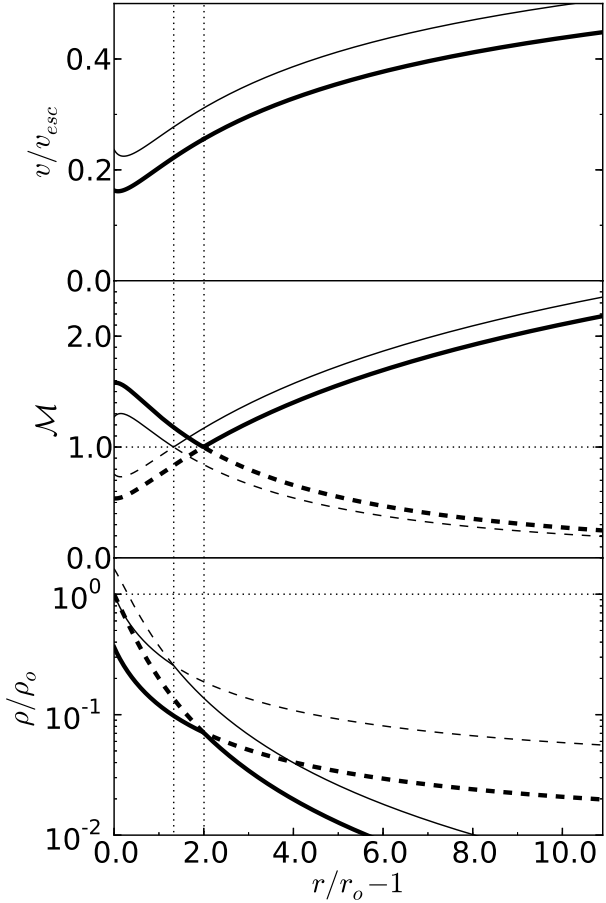}
	\vspace{-20pt}
	\caption{Top panel: bulk velocity profiles of the two transonic solutions corresponding to the two roots of \eqn{critpointeqn} for the parameter set chosen for a Parker wind with rotation by Keppens \& Goedbloed (1999): $\lp = 5.44995$, $\gamma = 1.13$, and $\zeta = 1.9$.  Middle and bottom panels: Mach number and density profiles of the four transonic solutions corresponding to the two roots.  The bold \textit{outflow} solution was displayed in Figure 1 of Keppens \& Goedbloed (1999); it has $r_c = 2.9894\, r_o$, $\lc = 7.6807$, $e_c = 5.9077$, $M_o = 0.5376$, $v_o = 0.1628\, v_{esc}$, and $v_\infty = 0.8770\,v_{esc}$.  Notice also that it displays a Mach number minimum at $r = 1.0360 \,r_o$. (The bulk velocity profile has a minimum at $r=1.0877\, r_o$.)  The second outflow solution has similar properties ($r_c = 2.3256 \,r_o$, $\lc = 6.5036$, $e_c = 5.7940$, $M_o = 0.7565$, $v_o = 0.2291\, v_{esc}$, and $v_\infty = 0.9439\, v_{esc}$) but does not satisfy the density boundary condition, having instead $\rho(r_o) = 1.628 \,\rho_o$.  }
\label{dblsolns}
\end{figure}

This second set of solutions does not arise if the boundary is at infinity, as in the Bondi problem, nor does it arise in the isothermal case because as $\gamma \rightarrow 1$, these two sets of solutions approach each other and coincide for $\gamma=1$.  In \apx{newcritpts}, we examine the properties of this second set of solutions for the classic Parker problem.  Suffice it to say here that \textit{without rotation}, only one of the two outflow solutions is a sought after wind solution satisfying $\rho(\x=0)/\rho_o = 1$.  

We should point out that Keppens \& Goedbloed (1999) alluded to the presence of multiple roots of the rotating Parker wind critical point equation, but they incorrectly identified one fast rotating solution as passing through two critical points.  To clarify our notion of two critical points, we reproduce their test solution in \fig{dblsolns}.  The top panel shows the two wind velocity profiles obtained for the two critical points; the bold outflow curve with a sonic point at $r_c \approx 3 r_o$ is the solution shown in Figure 1 of Keppens \& Goedbloed (1999).  As we explain in \sec{rotation}, what they called a second critical point is likely the location of the velocity minimum.    

In the middle panel, we show all four transonic solutions, separated into their subsonic (dashed) and supersonic (solid) branches. As shown in the bottom panel, only the bolded outflow curve satisfies the density boundary condition $\rho(r\rightarrow r_o) = \rho_o$, and so the other must be rejected.  The density boundary condition is also satisfied by the \textit{inflow} curve of the second set of transonic solutions.  It is tempting to suppose that this is always the case based on a symmetry argument: owing to the mathematical equivalence of the wind and accretion equations, why should the density boundary condition preferentially be satisfied by the outflow solution alone?  Indeed, this argument holds under spherical symmetry.  However, one of our more interesting results is that the inclusion of rotation can permit both outflow solutions to satisfy the density boundary condition.  How this comes about is addressed at length in \sec{edefrotation}, but we mention it here to draw attention to a very similar finding made by Cur\'e (2004), who reported that line-driven wind equations also have an `always present', second set of critical points when rotation is added to the problem.  As in our case, he showed that the second family of solutions does not ordinarily satisfy the density boundary condition but can at high enough rotation rates.  

\subsection{The Isothermal Solution} \label{LambertWfctsoln}
For an isothermal EoS, it has been pointed out that the entire problem can be compactly solved in terms of the Lambert $W$ function (Cranmer 2004). The many properties of the Lambert $W$ function and several of its uses in physics can be found in Valluri et al. (2000) and references therein.  An advantage of this solution method is that the Lambert $W$ function is built into Mathematica, Maple, and MATLAB.  We used it to expediently survey the parameter space of our isothermal disc wind solutions and to calculate the results presented in \sec{equivnozzle}.  

For $\gamma=1$, the equation of motion is reduced to
\beq F\equiv \bigp{1-\frac{1}{w}}w' + 2\bigp{g - \frac{\mathrm{A}'}{\mathrm{A}}} = 0 \label{F0iso}.\seq
Recalling that $g = dU_{eff}/d\x$, we obtain after integration
\beq \ln w -w = 2[U_{eff} - \ln (\mathrm{A}/\mathrm{A}_o)] + constant, \seq
where we have absorbed a factor of $\ln \mathrm{A}_o^{-2}$ into the constant.
Exponentiating and multiplying by $-1$ gives
\beq -w \exp(-w) = -\Gamma_B^2\bigp{ \f{\exp[U_{eff}]}{\mathrm{A}/\mathrm{A}_o} }^2 . \label{weweqn} \seq
Here, $\Gamma_B$ is a constant; its value is obtained by evaluating \eqn{weweqn} at the critical point, giving
\beq \Gamma_B = \f{\mathrm{A}_c}{\mathrm{A}_o}\exp\bigb{ -\f{1}{2} - U_{eff}(\x_c)}  .\label{gammaB}\seq
The magnitude of $\Gamma_B$ can be quite large ($\sim 10^2$) if the critical point is far away, since $\mathrm{A}_c/\mathrm{A}_o$ measures how much the flow has expanded before it becomes supersonic. 

`Operating' on both sides of \eqn{weweqn} with the Lambert $W$ function isolates $-w$.  Hence, the solution in terms of the Mach number $\mathcal{M} = \sqrt{w}$ is
\beq \mathcal{M}(\x) =\sqrt{ -W\bigb{- \bigp{\Lambda \Gamma_B \f{\exp[U_{eff}(\x)]}{\mathrm{A}(\x)/\mathrm{A}_o }}^2 } }. \label{Wfctsoln} \seq
We have re-introduced $\Lambda$ into \eqn{Wfctsoln} to distinguish the transonic $(\Lambda =1)$ solutions from the everywhere sub/supersonic solutions ($\Lambda < 1$).  Here again, setting $\Lambda > 1$ yields double-valued solutions.

The location of the critical point is obtained directly from \eqn{criteqn}; $\x_c$ must satisfy
\beq g_c = \f{\mathrm{A}'_c}{\mathrm{A}_c}. \label{isosingcdn} \seq
For our disc wind models, $g_c$ is given by \eqn{geff} and $\mathrm{A}'_c/\mathrm{A}_c = q\cos i/( f  + \x_c \cos i)$, where $q=1$ for the CIA and model and $q=2$ for the Converging model.  Due to the non-linear dependence on $\cos i$, $\x_c$ must again be solved for numerically.  For the spherically symmetric Parker model, meanwhile, $\mathrm{A}'_c/\mathrm{A}_c = 2/( f  + \x) = 2/(r/r_g)$ (recall that $r_g = \lp r_o$), while the gravitational force is simply $g = 1/(r/r_g)^2$.   We immediately recover the well known sonic point distance $r_c = r_g/2$ from \eqn{isosingcdn}.  The cylindrical Parker wind model (i.e. the CIA model at $i=0^\circ$) discussed in \sec{CylindricalParker} also has $g = 1/(r/r_g)^2$, while $\mathrm{A}'_c/\mathrm{A}_c =  1/(r/r_g)$, giving $r_c = r_g$.  We can therefore in general expect the CIA model to have critical points approximately twice as distant from those of the Converging model.     

The mass loss rate is calculated from $d\dot{M} = \dot{m}A_o$ by knowing the mass flux density, $\dot{m}= \rho_o c_o \mathcal{M}_o$.  From \eqn{Wfctsoln}, we find that
\beq \mathcal{M}_o =\sqrt{ -W\bigb{-(\Lambda \Gamma_B)^2 \exp(-2\lp + \zeta^2) }}. \label{Moiso} \seq
By the definition of the Lambert W function, we can instead express \eqn{Moiso} as the nonlinear relationship
\beq  \mathcal{M}_o=\Lambda\Gamma_B\exp{\bigb{-\lp + \f{\zeta^2 }{2} + \f{\mathcal{M}_o^2}{2}}} . \label{mdotiso}\seq
To an excellent approximation when $\mathcal{M}_o << 1$ (valid when $\lp >> 1$), therefore, the mass flux density is given by
\beq \dot{m} = \rho_o c_o\Lambda\Gamma_B\exp{\bigb{-\lp + \f{\zeta^2 }{2}}} .\label{mdotiso2}\seq

The density distribution follows immediately from the continuity equation, $\dot{m} = c_o \rho \mathcal{M}\mathrm{A}/\mathrm{A}_o$:
\beq \f{\rho(\x)}{\rho_o} = \f{\mathcal{M}_o/\mathcal{M}(\x)}{\mathrm{A}(\x)/\mathrm{A}_o} = \f{\Lambda \Gamma_B\exp\bigb{-\lp + \zeta^2/2 + \mathcal{M}_o^2/2}}{(\mathrm{A}(\x)/\mathrm{A}_o)\mathcal{M}(\x)} ,\label{rhoiso}\seq
where we used \eqn{mdotiso} to obtain the second equality.
It should be verified that $\rho(\x=0)/\rho_o = 1$ upon implementation of these equations.    

The barometric law that we quoted in \sec{hydrostatics} is derived from \eqn{rhoiso}.  We consider a spherically symmetric ($\zeta = 0$) Parker wind applied to a isothermal planetary atmosphere.  Taking $r_o$ to be the radius of the exobase, $\rho_o$ is the density at this height.  Such an atmosphere can be modeled as a transonic Parker wind if, in the steady state, the top level of the atmosphere is itself moving at speeds approaching the speed of sound at the exobase.  On the other hand, a static atmosphere, undergoing mass loss via evaporation, should approximately resemble a Parker wind solution with $\dot{m} << \dot{m_c}$ (corresponding to an everywhere subsonic solution with a finite density at infinity).  
Looking to \eqn{Wfctsoln}, we see that for small mass-loss rates ($\Lambda << 1$), the argument of $W$ is small and it is valid to expand $W$ to leading order.\footnote{$W(x) \approx x -x^2 - \f{3}{2}x^3 + \dots$}  Thus, \eqn{rhoiso} becomes
\beq \rho= \rho_o\exp\bigb{\lp(r_o/r) -\lp + \mathcal{M}_o^2/2} .\label{barometriclaw}\seq
For a strictly static atmosphere ($\mathcal{M}_o = 0$), we recover a barometric law, but one derived from hydrodynamic (rather than hydrostatic) equilibrium.  

%===============================================================================
% RESULTS
%===============================================================================

\section{Results}\label{results}

\subsection{Estimating $\dot{M}$ using 1D Models} \label{polymdot}

The scaling relationships of isothermal Parker wind solutions have been used to predict global mass loss rates from protoplanetary discs (Adams et al. 2004; Gorti \& Hollenbach 2009).  For instance, Gorti \& Hollenbach (2009) combine a realistic treatment of the photevaporative heating process in protoplanetary discs by separately calculating the isothermal EUV ionization front and the FUV and X-ray heated neutral flow surface, thereby self-consistently determining the base density and temperature on the flow boundary.  Since their radiative transfer calculation assumed hydrostatic equilibrium ($\mathcal{M}_o = 0$), the hydrodynamics had to be included separately to calculate the mass loss rate.  The latter is obtained by integrating  the mass flux density, $\dot{m} = \rho_o v_o = \rho_o c_o \mathcal{M}_o$, via $\dot{M} = 4\pi \int_{r_{in}}^{r_{out}} \dot{m} \sin i\, r_o \, dr_o$.  In other words, incorporating the hydrodynamics `by hand' given $c_o$ and $\rho_o$ requires determining a nonzero value for $\mathcal{M}_o$.  Employing isothermal Parker wind solutions offer a convenient way to accomplish this, for the HEP is obtained from $c_o$ and the solutions are independent of $\rho_o$.  Given a value of the HEP, the sonic point distance is uniquely specified by \eqn{isosingcdn} and the corresponding initial Mach number is obtained from \eqn{Moiso}.

\subsubsection{Isothermal Scaling Relationships} \label{isomdot}
Using their numerical results for $\rho_o$ and $c_o$, Gorti \& Hollenbach (2009) obtained the mass flux density as a function of $r_o$ by utilizing the scaling relationship for $\dot{m}$ derived by Adams et al. (2004).  This scaling relationship is more generally our \eqn{mdotiso2} with Keplerian rotation ($\zeta = \sqrt{\lp}$) assigned and assuming transonic outflows ($\Lambda=1$):
\beq \dot{m} = \rho_o c_o \Gamma_B \exp(-\lp/2). \label{Moisoapx} \seq
The quantity $\Gamma_B$ depends on the specific streamline geometry and is given by \eqn{gammaB}.  
The term $\exp{(-\lp/2)}$ controls the mass flux density for large HEP.  The exponential dependence results from the logarithmic enthalpy term and acts to suppress the wind whenever when the thermal energy of the gas is small compared to the escape velocity, e.g. when the gas is deep in the potential well of the inner disc region and shielded from high energy photons.   

As we mentioned in \sec{heating}, BMS83 found the same exponential dependence in their isothermal wind region E, namely $\dot{m} \propto \exp{(-T_g/2T_{IC})}$, where $T_g$ is the `escape temperature' defined by $kT_g = \mu m_p V_{esc}^2$.  To make the comparison explicit, we must recall \eqn{HEPbms83}, which says for $\gamma=1$ and $\xi = r_o/R_{IC} = T_{IC}/T_g$ that the HEP is simply $\lp = T_g/T_o$.
For a tightly bound corona heated to the Compton temperature, we have that $T_o = T_{IC} < T_g$, so we can indeed identify the HEP as being equal to $T_g/T_{IC}$.

\subsubsection{Polytropic Scaling Relationships} \label{polymdot}
Similar agreement can be found using polytropic models, which are able to sample a larger range of thermodynamic conditions and can therefore lead to more accurate disc dispersal time-scale estimates.  The procedure for incorporating the hydrodynamics `by hand' using Parker wind or Parker-like disc wind models is the same as that given above, the only change being that the initial Mach number is now given by $\mathcal{M}_o =(\mathrm{A}_c/\mathrm{A}_o)(\lp/\lc)^\f{\gpone}{2(\gmone)}$.  (Correspondence with the isothermal result can be obtained using $\rho_c/\rho_o = (\lp/\lc)^{1/(\gamma-1)}$.)  Thus, for polytropic winds, the exponential terms in \eqn{Moisoapx} are replaced by a strong functional dependence on the temperature at the sonic point ($\lp/\lc = T_c/T_o$), so the mass flux density scales as 
\beq \dot{m} =\rho_o c_o (\mathrm{A}_c/\mathrm{A}_o)\bigp{\f{T_c}{T_o}}^\f{\gpone}{2(\gmone)} .\seq
For $\gamma = 5/3$, we recover the temperature dependence found by BMS83 in their Region C, namely $\dot{m} \propto (T_c/T_o)^2$.  (To make the comparison with BMS83, $T_c$ is to be associated with their `characteristic' temperature $T_{ch}$ and $T_o$ with $T_g$, obeying $T_{ch} < T_g < T_{IC}$.)

By \eqn{criteqn}, we can instead express the initial Mach number as
\beq \mathcal{M}_o = \f{\mathrm{A}_c}{\mathrm{A}_o}\bigp{g_c\f{\mathrm{A}_c}{\mathrm{A}'_c}}^\f{\gpone}{2(\gmone)}. \seq
We see that $\mathcal{M}_o$ now depends sensitively on the effective gravitational force instead of $\exp{(-U_{eff})}$, as well as on the ratio of the flow tube area and the streamline divergence at the critical point.    

\subsection{Isothermal Flow Properties using Nozzle Functions} \label{equivnozzle}

\begin{figure}
\includegraphics[scale=0.4]{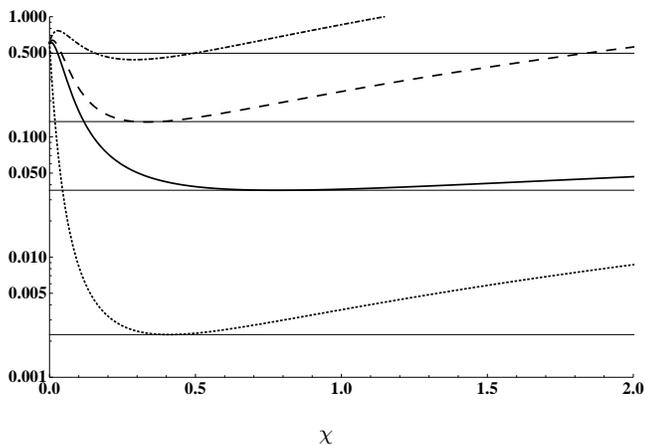}
\begin{center} {$\x$} \end{center}
\caption{Equivalent nozzle functions $N(\x)$ for a Keplerian Parker wind (topmost dashed-dotted curve), the $i=60^\circ$ Converging (dashed) and CIA (solid) models, and the spherically symmetric Parker wind (bottom dotted curve).  Rotating these nozzle functions about the $\x$-axis sweeps out the area of the de Laval Nozzle having steady-state flow properties identical to that of the wind.  $N(\x)$ is normalized so that $N(\x_c) =\Gamma_B \exp(-\lp/2)$, which is approximately $\mathcal{M}_o$.  The horizontal lines give the exact value of $\mathcal{M}_o$, calculated using \eqn{Moiso}.  They are, from top to bottom, $\mathcal{M}_o=[ 0.497,0.134,0.036,0.002]$, with corresponding critical points (throat positions) $\x_c =[ 0.290, 0.333,0.779,0.409]$.  All nozzle functions were calculated with $\lp =11$.  The hump on the topmost curve is a bulk velocity minimum, located at $\x = 0.029$.}
\label{nozzleplot}
\end{figure}

A useful means for gauging how the flow properties (i.e. sonic point distance, initial Mach number, mass flux density, and acceleration) of our disc wind models compare to those of Parker winds with or without Keplerian rotation is obtained by defining the equivalent nozzle function (see Parker 1963 for the polytropic nozzle function).
The well-known equation of motion for the isothermal de Laval nozzle is, in our notation,
\beq  \bigp{1-\frac{1}{w}}w' = 2\frac{N'}{N}  ,\label{laval}\seq 
where $N = N(\x)$ is the cross-sectional area of the nozzle at any distance $\x$.  
Making the comparison with \eqn{F0iso}, $N(\x)$ is obtained by solving $d\ln N = d\ln A - dU_{eff}$, giving
\beq N(\x) = \f{A(\x)}{A_o}\exp \bigb{-U_{eff}(\x) - \bigp{\lp - \f{\zeta^2}{2} + \f{1}{2}}} .\seq
We have normalized $N(\x)$ so that $N(\x_c) =\Gamma_B \exp(-\lp/2)$, which is the leading order approximation to $\mathcal{M}_o$ via an expansion of \eqn{Moiso}.  Since the critical point occurs at the `throat' of the nozzle, where the cross-sectional area is a minimum, $N(\x)$ is a visual tool that can be used to find both the sonic point and the approximate initial Mach number (and hence mass-flux density) by inspection.  In \fig{nozzleplot}, we plot the nozzle functions of the Keplerian rotating Parker wind (dashed-dotted line),  the Converging Model at $60^\circ$ (long-dashed line), the CIA model at $60^\circ$ (solid line), and the spherically symmetric Parker wind (dotted line) for $\lp = 11$.  The horizontal lines correspond to the exact value of the initial Mach number from \eqn{Moiso}.  We see that all of the horizontal lines except that of the Keplerian Parker wind with $\mathcal{M}_o \approx 0.5$ intersect very near the minimums of the nozzle functions, showing that \eqn{Moisoapx} is an excellent approximation when $\mathcal{M}_o$ is small.   

A noticeable feature of the nozzle functions is the initial hump close to the opening (near $\x=0$) for the models undergoing Keplerian rotation.  The spherically symmetric Parker wind nozzle is everywhere converging before the throat and diverging thereafter, thus ensuring that the flow will never decelerate.  The presence of the humps indicates that the flow is entering a diverging nozzle ($N'> 0$), so that for initially subsonic flow, we must have $w'< 0$ by \eqn{laval}: the flow decelerates until reaching the top of the hump where $N'= 0$.  The flow is still subsonic at this location, implying that the acceleration must be zero ($w' = 0$), i.e. the flow has reached its minimum velocity.  The flow then proceeds to accelerate with the converging nozzle, traverse the sonic point at the throat where $w = 1$ and $N' = 0$ (but $w' \neq 0$), and continues to accelerate supersonically ($w'>0$ and $w>1$) in the diverging region where $N' > 0$. 

Comparing the nozzle functions of the CIA and Converging models, it is clear that the former model has a sonic point about twice as distant as the latter, as expected.  More distant sonic points imply smaller initial Mach numbers for a given HEP, and since $\mathcal{M}_o$ is a direct gauge of the mass flux density, the total mass loss rate for a CIA wind will also be smaller in general.  These differences all result from the halted expansion room of the CIA model, as will become clear in \sec{psurvey}.  Both winds experience a reduced centrifugal force at $i = 60^\circ$, explaining why the Keplerian Parker wind has a significantly higher initial Mach number.  We can therefore arrive at the result that the mass flux densities of our disc wind models are always bounded from below by that of the spherically symmetric Parker wind and above by that of the Keplerian Parker wind.  

More generally, plotting $N(\x)$ allows one to easily infer the effects of altering the geometry of the flow or the effective potential.  For instance, the humps practically disappear by setting $A(\x)=A_o$.  Recalling \fig{stefans}, the streamlines found by Luketic et al. (2010) first originate from the disc midplane in a more vertical fashion before bending radially, implying that the area between streamlines indeed behaves as if $A(\x)=A_o$ for very small $\x$.  Therefore, it is likely that the velocity minimums would not occur in a model that captures this feature, although it is worth noting that Luketic et al. (2010) observe non-monotonic radial velocity profiles in their fiducial run (see their Figure 5). 

\subsection{Comparison with Hydrodynamical Simulations}  
Having found an analytical wind solution for a geometry that approximates the simulation results in \fig{stefans}, we now attempt to reproduce the shape of the sonic surface.  
Luketic et al. (2010) reported that in the region of self-similar flow, this shape is approximately a straight line given by $z_c = a x$ (in units of $R_{IC}$), with a slope $a \approx 1/4$ for their fiducial run (see their Figure 1) and $a \approx 1/3$ for their isothermal run (unpublished).  Similar shapes and slopes can be seen in several of the figures of Woods et al. (1996) and Font et al. (2004).  From \fig{modelgeometry2}, we can equivalently specify a linear sonic surface as 
\beq \f{l_c}{r_o} = \bigp{\f{\sin i}{a} - \cos i }^{-1} .\seq
By definition, the sonic point of our analytic solution is the location satisfying
\beq \f{l_c}{r_o} = \lp \x_c .\seq
Therefore, a qualitative comparison can be made by assuming that the base of the wind has a constant HEP over the emitting region.
Physically, this corresponds to a temperature profile satisfying $T_o \propto r_o^{-1}$ on the midplane.  We see that the CIA model can accurately reproduce the sonic surface found by Luketic et al. (2010) if $\lp \x_c$ can be as small as $\bigp{\sin i/a - \cos i }^{-1}$ (which for $i=30^\circ$ is about $0.88$ for $a=1/4$ and $1.58$ for $a=1/3$).  Unfortunately, the product $\lp \x_c$ is always found to be greater than 2 for $1\leq \gamma \leq 5/3$, showing that multidimensional effects cannot be ignored.  Nevertheless, the sonic surface of the CIA model will still closely resemble that of the simulation results, having slope of about $40^\circ$ for the smallest values of $\lp \x_c$.  

In order to make a more quantitative comparison with simulations, it would be necessary to know the variation of the HEP as a function of $r_o$.  A detailed comparison along these lines will be presented in a follow up paper.  

\subsection{Parameter Survey of Polytropic Transonic Disc Wind Solutions}\label{psurvey}
\begin{figure*}
	\centering
	\includegraphics[scale=0.5]{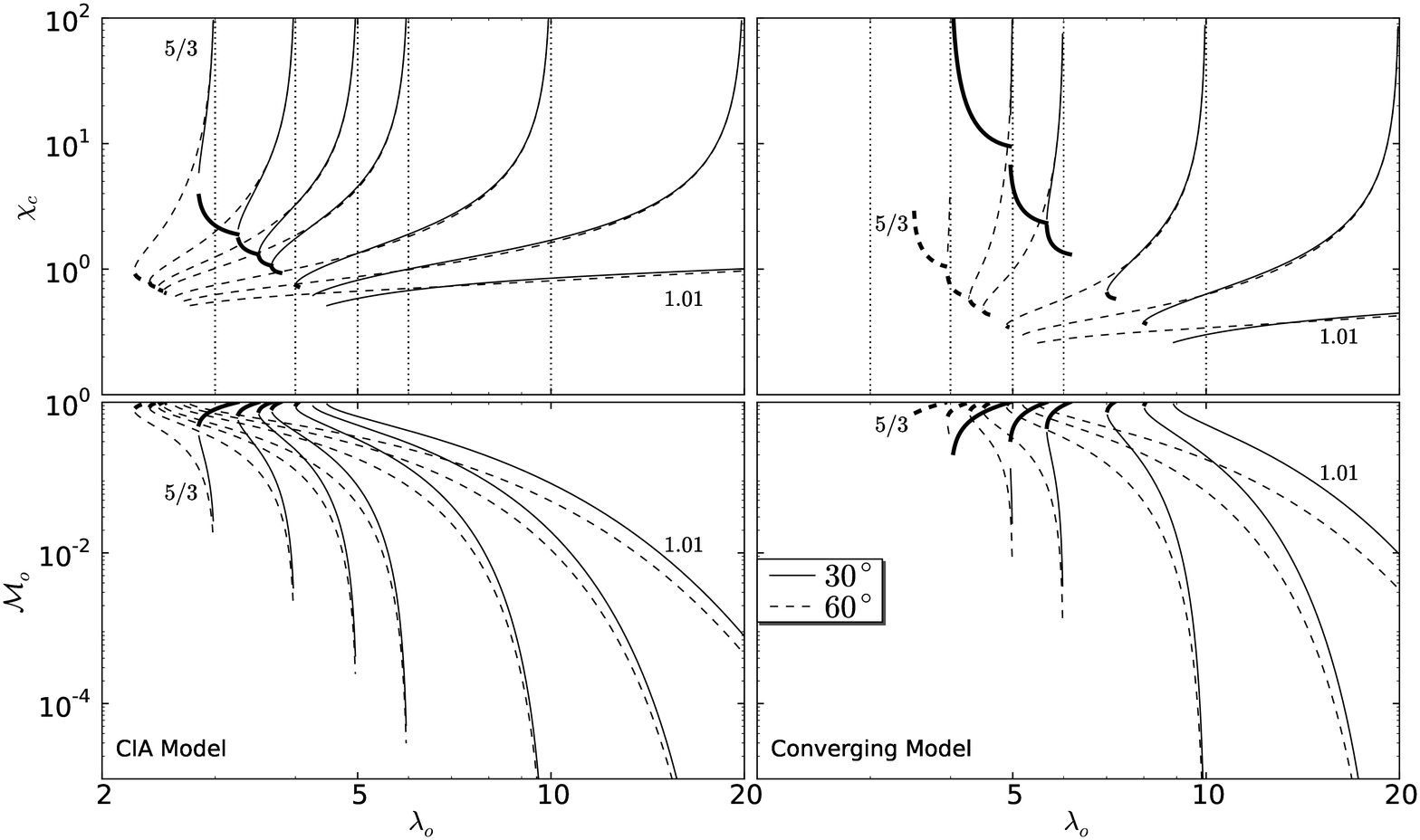}
	\caption{Parameter survey for the CIA and Converging models for two inclination angles, $i=30^\circ$ and $i = 60^\circ$.  Each curve has a constant $\gamma$; the polytropic indices between $\gamma = 1.01$ and $\gamma = 5/3$ are $\gamma = (1.1, 1.2,4/3,1.4,\&1.5)$.  Initial Mach numbers in the lower panels correspond to the sonic point distances in the upper panels for the same $\lp$.  The `tails' of critical point curves with $\x_c$ decreasing ($\mathcal{M}_o$ increasing) with increasing $\lp$ are shown in bold.  Vertical dotted lines at $\lp=(3,4,5,6,7)$ mark the value $\lp=2/(\gamma-1)$; critical points to the right of this line correspond to transonic solutions with $v_o > v_\infty$.}
	\label{dwcritpts}
\end{figure*}

The parameter space of our disc wind models is all values of $(\gamma, \lp, i)$ that lead to transonic solutions.  For a fixed $\gamma$ and $i$, a parameter survey is encapsulated by a plot of the critical point distance as a function of HEP, which we will refer to as a critical point curve.  Limiting our attention to the two intermediate angles $i=30^\circ$ and $i=60^\circ$, we display in the top panels of \fig{dwcritpts} critical point curves for select $\gamma$ ranging from nearly isothermal ($\gamma = 1.01$) to adiabatic ($\gamma = 5/3$).  Corresponding initial Mach numbers are given in the bottom panels.  

The shape of the $\gamma=1.01$ critical point curves can be qualitatively understood by considering the isothermal Parker model, in which $r_c/r_g = \x_c + 1/\lp = 1/2$.  For large HEP, the centrifugal term in the effective potential is small, and the Converging model will closely resemble the Parker model.  Indeed, its critical point curve is approximately $\x_c= 1/2 - 1/\lp$, with the $1/\lp$ explaining the decreasing trend in $\x$ as $\lp$ decreases.  Meanwhile, we pointed out previously that the isothermal CIA model has sonic points about twice as distant as the Converging model, which can be seen from the top left panel. 

Transonic solutions at higher $\gamma$ require progressively smaller HEP values (e.g., higher temperatures) to make up for the energy lost to $PdV$ work.  Focusing on a fixed HEP value, e.g. $\lp = 6$, the critical points are shifted to ever larger distances as $\gamma$ increases.  This occurs despite the fact that the gas will cool faster for larger $\gamma$ (thus lowering the sonic threshold to smaller bulk velocities) because the launching velocity becomes substantially smaller at larger $\gamma$.  Transonic solutions with the least distant critical points have very high initial Mach numbers and delimit the edge of the parameter space.  The dotted vertical lines that bound the HEP for large $\x_c$ -- regardless of model and inclination angle -- have the value $2/(\gamma-1)$.  For a given HEP in either model, the $i=60^\circ$ solutions are launched with smaller initial Mach numbers than the $i=30^\circ$ solutions.  This is a simple consequence of the reduction in centrifugal force at larger inclination angles.  The extra rotational energy raises $\mathcal{M}_o$ for $i=30^\circ$ above $\mathcal{M}_o$ for $i=60^\circ$.   Conversely, the reduced centrifugal force for $i=60^\circ$ permits critical points to extend to smaller HEP before reaching $\mathcal{M}_o \approx 1$.  

A peculiar feature of \fig{dwcritpts} is the appearance of a `tail' on the critical point curves (shown in bold), signifying that at the lowest HEP for any given curve with $\gamma \ga1.1$ ($\ga 1.2$ for the CIA model), there are two transonic solutions.  As we mentioned in \sec{topology}, this occurs when the outflow solution corresponding to the second root of the critical point equation satisfies the density boundary condition.  The tail of the critical point curve is generally in close proximity to the disc, so these transonic solutions have a higher initial Mach number than the `normal' solutions.  As $\gamma$ increases, this tail grows in length, while the normal critical point curve shrinks.  Only the tail extends to the right of the vertical lines at $\lp = 2/(\gamma -1)$; solutions on this side of the line have the property that their initial velocity exceeds their terminal velocity.  Notice that the normal critical point curve has an increasing slope, and one would intuitively expect the critical point to shift downstream as the temperature is decreased.  Meanwhile, the tail displays the opposite behavior, so that more distant sonic points with lower initial Mach numbers have \emph{higher} temperatures.  We postpone an explanation of this behavior to \sec{discussion}.

Our parameter survey reveals that disc winds possess solutions for $\gamma = 5/3$, in contrast to Parker winds (with or without rotation).  The top right panel shows that the Converging model has tail solutions to the right of the vertical line at $\lp=3$ for a narrow range of HEP ($3.5 \la \lp \la 4.0$) for $i = 60^\circ$.  The bottom panel reveals that these solutions have initial Mach numbers close to unity.  There are no solutions for the Converging model at $i = 30^\circ$ for $\gamma = 5/3$ (at least not within $l=10^2 r_g$).  Meanwhile, the CIA model has both normal and tail solutions at both angles, the normal ones permitting small initial Mach numbers.  

Overall, this parameter survey indicates that the CIA model, with its lack of streamline divergence, gives rise to what appears to be a more versatile wind.
At small $\gamma$, for instance, the CIA model has transonic solutions that begin for $\lp$ about half as small as the minimum HEP allowed for the Converging model.  At larger $\gamma$, the critical points for the CIA model span a more appreciable range of HEP and altogether dominate for $\gamma=5/3$.  

These properties, which are solely due to geometric differences, are physically a manifestation of the rate of enthalpy dissipation.  Just considering the Bernoulli function, it is clear that the more rapidly that heat is liberated, the faster the flow must become to keep $B_o$ constant.  Converging streamlines exhibit both lateral expansion due to streamline divergence and azimuthal expansion as the wind cone widens, so the enthalpy can dissipate faster than it can with the CIA model.  Conversely, the confined expansion imposed by the CIA streamline configuration allows the flow to retain more of its enthalpy as it expands, so that the flow can be launched at smaller $\lp$ before the initial Mach number approaches unity.  As $\gamma$ increases and the flow starts off with less enthalpy, a smaller compensatory reduction in HEP is required to launch transonic solutions compared to the Converging model.

\subsection{Disc Wind Acceleration Zones}
Each solution corresponding to the critical point locations in \fig{dwcritpts} has an associated acceleration zone, $l_{90}$, which we define as the distance where the flow reaches $90\%$ of its terminal velocity, i.e. $v(l_{90})/V_{esc} = 0.9 \sqrt{2 e_c/\lc}$ (see \eqn{vterminal}).  A nearly isothermal wind has its temperature held constant to very large distances, so the acceleration zone extends far beyond a closer to adiabatic wind whose temperature falls off rapidly.  For a given $\gamma$, the acceleration zone is also sensitive to the flow geometry.  The CIA model has a greatly extended acceleration zone, again attributable to its geometrical confinement curtailing adiabatic expansion.  For example, $l_{90}\sim10^6 \,r_g$ for $\gamma=1.1$ compared to $10^4 \,r_g$ for the Converging model.  Apparently, the factor of two separation in the critical point distance between these models translates into a two order of magnitude difference in the extent of the acceleration zone!  As $\gamma$ increases, the acceleration zone moves progressively closer to the equatorial plane, but there remains a two order of magnitude separation between our disc wind models.
 
\subsection{The CIA vs. the Converging Model} \label{ciavscon}

\begin{figure}
	%\centering
	\includegraphics[scale=0.55]{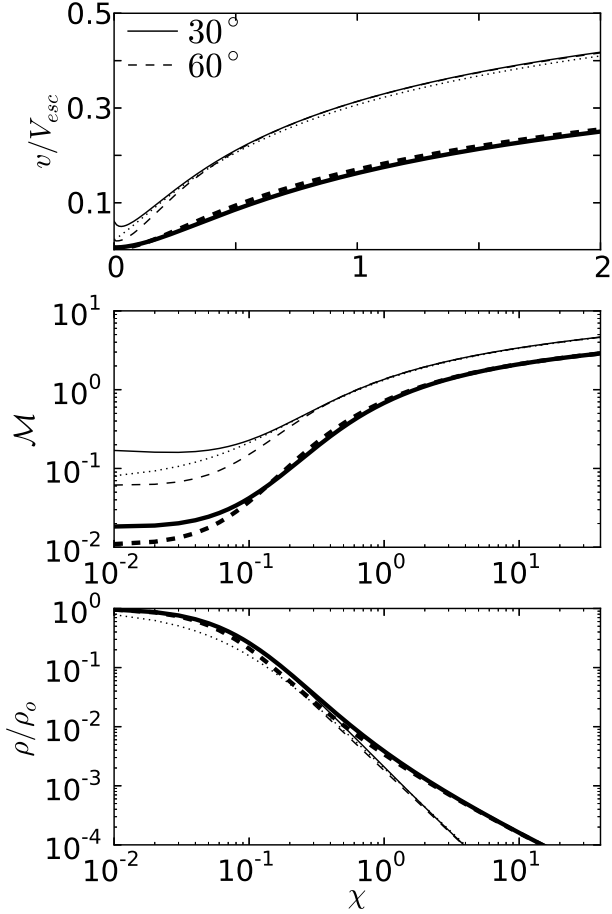}
	\vspace{-20pt}
	\caption{Transonic disc wind solutions for the Converging model (unbolded) and CIA model (bolded) with $\gamma = 1.1$ and $\lp=10$.  Also shown is a $\gamma=1.1$, $\lp=5$ Parker wind solution (dotted; the reduction in HEP is equivalent to making $v_{esc} = V_{esc}$).  Top panel: bulk velocity profiles for the first $2 l/r_g$.  Middle Panel: Mach number profiles.  Bottom panel: Density profiles.  Properties of these solutions are given in \tbl{2}. }
\label{dwplot1}
\end{figure}

\begin{table*}
\small
\begin{center}
\caption{Properties of the $\gamma=1.1$ transonic solutions plotted in \fig{dwplot1}:}
\begin{tabular}{l c c c c c c c}  \\ \hline
\hline 
Model, $i$ & $\lp$ & $\x_c$ & $\lc$ & $e_c$  & $\mathcal{M}_o$ &$v_o/V_{esc}$ & $v_\infty/V_{esc}$  \\
\hline
 Parker & 5. &0.6496 & 8.4960 & 8.5000 & 0.0690 & 0.0218 & 1.0002 \\
 Con, $30^\circ$& 10. &0.6327 & 16.7032 & 8.3825 & 0.1922 & 0.0608 & 1.0018 \\
 Con, $60^\circ$& 10. &0.6270 & 17.0044 & 8.5058 & 0.0649 & 0.0205 & 1.0002  \\
 CIA, $30^\circ$& 10. &1.7003 & 18.9552 & 9.4780 & 0.0191 & 0.0060 & 1.0000  \\
 CIA, $60^\circ$& 10. &1.6270 & 18.9680 & 9.4841 & 0.0110 & 0.0035 & 1.0000 \\  
\hline \hline 
\end{tabular}
 \end{center}
\normalsize
\label{dwtable1}
\end{table*}

Here we compare transonic solutions for our two streamline geometries.  Recalling our \fig{modelgeometry2}, the 2D disc wind model that we are attempting to explore with our 1D solutions is comprised of two wind regions.  The inner region hosts Converging streamlines beginning at $i\approx 60^\circ$, that then diverge out to some distance along the midplane until the inclination angle coincides with that of the outer wind region, occupied by CIA streamlines at $i \approx 30^\circ$.  

We limit our comparison to plotting transonic solutions for the same set of parameters $(\lp,\gamma)$ for either model with angles $i= 30^\circ$ and $i= 60^\circ$.  In \fig{dwplot1} we show the bulk velocity profiles within $2l/r_g$ (upper panel), as well as the Mach number and density profiles on a larger scale (lower panels range to $\x=40$), for parameters $\lp = 10$ and $\gamma = 1.1$.  The properties of these solutions are tabulated in \tbl{2}.  We note first off that the inclination angle only affects the transonic solutions in the subsonic flow regions.  Secondly, there is an order of magnitude difference in the initial velocities of the CIA and Converging models.  This has the following implication for kinematic models, which typically assume a velocity law based solely upon $v_o$, $v_\infty$, and a parameter controlling the slope of the velocity: the values of $v_o$ and the velocity gradient are quite sensitive to the type of wind geometry.  Winds launched from $i=30^\circ$ in either model have higher initial velocities due to the increased centrifugal force, in agreement with \fig{dwcritpts}.  Despite having substantially different values of $v_o$, all four solutions tend to nearly the same terminal velocity because the Bernoulli constant $B_o = v_\infty^2/2$ should be unchanged for a similar set of wind parameters.  Note that this is not obvious from \fig{dwplot1} because the CIA model, with its extended acceleration zone, is yet to reach its terminal velocity at $\x = 40$.  
 
Aside from the appearance of velocity minimums on the bulk velocity profiles, the Converging wind is well approximated by a spherical Parker wind with an HEP half as great.  (Since the gravitational binding energy is halved for gas in a Keplerian disc, a disc wind HEP twice that of a Parker wind preserves the ratio of the escape velocity to the initial sound speed.)  Meanwhile, the CIA model gives rise to a significantly slower wind.  We can quantify the acceleration in the subsonic region using simple kinematics.  The average value of the acceleration between the midplane and the sonic point is $\avg{a} = (v_c^2 - v_o^2)/2l_c$.  Noting that $v_c = c_s(l_c)$, we have in terms of our tabulated quantities,
\beq \f{\avg{a}}{c_o^2/r_o} = \f{1/\lc - \bigp{v_o/V_{esc}}^2}{2\x_c} \label{aavg}, \seq 
from which we find that $\avg{a}_{CON} \approx 2.9 \avg{a}_{CIA}$ for the solutions in \fig{dwplot1}.  Equation \eqref{aavg} is also useful for comparing winds that undergo significant deceleration upon leaving the midplane (which occurs for large $\gamma$), in which case $\avg{a}$ will be negative.     

Although not noticeable, the bulk velocity profiles for the CIA model also have minimums close to the midplane.  
Gas rotating at Keplerian speeds \emph{must} initially decelerate upon leaving the equatorial plane, independent of the flow parameters.  This can be seen from \eqn{F0}; evaluated at $\x=0$ for $\zeta^2 = \lp$, we have that $y'<0$ provided $q>0$ (and $q=1$ for the CIA model and 2 for the Converging model). 
This situation results from the balance between centrifugal and gravitational forces on the disc midplane, so that the streamline divergence ($d\ln A/dl$) controls whether or not the flow can accelerate.  The flow will decelerate so long as $d\ln A/dl > 0$, which is always the case with streamlines that are straight in the $(x,z)$-pane, for otherwise they would have to intersect. 

The bottom panel of \fig{dwplot1} shows that the density of the CIA model varies asymptotically as $\rho \propto l^{-1}$, whereas the Converging model has $\rho \propto l^{-2}$.  This is a simple consequence of the continuity equation, the CIA model having an area term $A\propto l$ asymptotically.  We see that the density at the critical point is smaller in the CIA model because the CIA wind remains subsonic out to distances nearly three times as large as the Converging wind; the flow has a larger distance over which to expand.  This reduced acceleration also accounts for why the CIA model has a larger critical point constant and a smaller temperature at the critical point (recall that $\lc/\lp = T_c/T_o$) -- see \tbl{2}.       

\subsection{A Comparison of Degenerate Transonic Solutions}

\begin{table*}
\small
\begin{center}
\caption{Properties of the degenerate $\gamma = 5/3$ transonic solutions for the CIA model plotted in \fig{dwplot3}:}
\begin{tabular}{l c r r c r c c c}  \\ \hline
\hline 
$i$, (root) & $\lp$ & $\x_c$ & $\lc$ & $e_c$  & $\mathcal{M}_o$ &$v_o/V_{esc}$ & $v_\infty/V_{esc}$ \\
\hline
$30^\circ$, (tail) & 2.9 & 2.6585 & 9.7132 & 0.9517 & 0.6843 & 0.4018 & 0.4427  \\
$30^\circ$, (normal) & 2.9 & 15.3601 & 46.4707 & 0.9915 & 0.1541 & 0.0905 & 0.2066 \\
$60^\circ$, (tail) & 2.3 & 0.8292 & 3.3360 & 1.1330 & 0.9286 & 0.6123 & 0.8242  \\
$60^\circ$, (normal) & 2.3 &1.2653 & 4.5759 & 1.0790 & 0.6203 & 0.4090 & 0.6867  \\
\hline \hline 
\end{tabular}
 \end{center}
\normalsize
\label{dwtable3}
\end{table*}

\begin{figure}
	%\centering
	\includegraphics[scale=0.55]{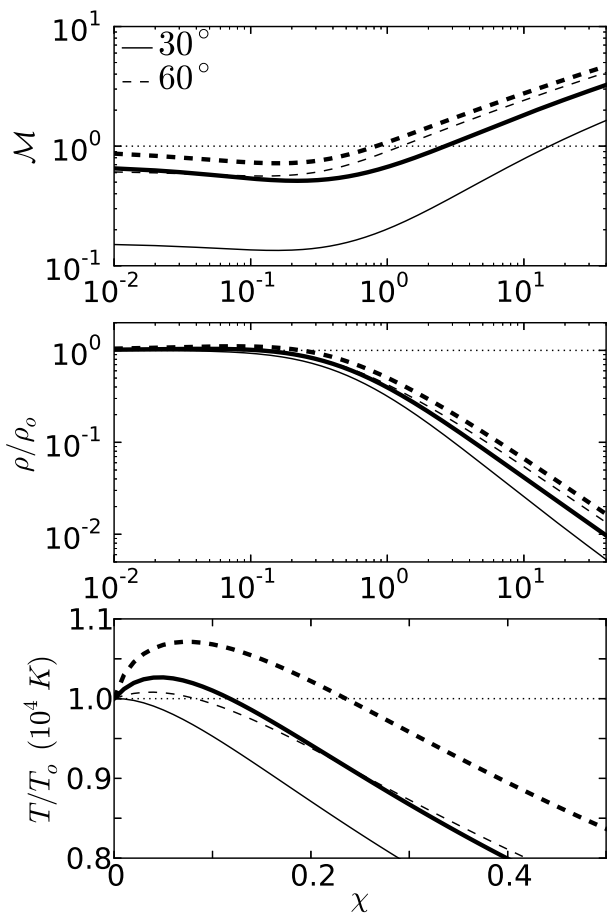}
	\vspace{-20pt}
	\caption{Degenerate transonic disc wind solutions for the CIA model with $\gamma = 5/3$; bolded solutions lie on the tail of the $\gamma=5/3$ critical point curves in \fig{dwcritpts}.  These tail solutions have substantially different properties -- see \tbl{3}.  The bottom panel displays the behavior of the temperature within $r_g/2$: the highest velocity solutions undergo a dramatic increase in temperature in the region of deceleration (i.e. the gas is adiabatically compressed). }
\label{dwplot3}
\end{figure}

Referring to to \fig{dwplot3}, we now examine two pairs of degenerate solutions for the CIA model with $\gamma = 5/3$: one pair with $\lp = 2.3$ for $i=60^\circ$ and another with $\lp = 2.9$ for $i=30^\circ$ -- see \tbl{3}.  From the Mach number profiles in the top panel, it is evident that each of these winds undergo marked deceleration before becoming sonic.  Indeed, $\avg{a}$ is slightly less than zero for the tail solutions and only slightly greater than zero for the normal solutions.   

In the bottom panel we zoom in on the region of the middle panel with $\x \leq 0.5$, in order to show the behavior of the temperature ($T(\x)/T_o = (\rho(\x)/\rho_o)^{\gamma -1}$) just above the midplane.  The temperature is shown in units appropriate for a photoionized disc heated to $\sim 10^4$ K; it is clear that the gas undergoes substantial adiabatic compression upon rising above the midplane, being heated by as much as $700K$ for the $i = 60^\circ$ tail solution.  This behavior epitomizes the dilemma posed by degenerate solutions: if dust formation is to be taken into account, which of the two profiles are we to believe?  

The heating is not unique to $\gamma = 5/3$ and can be explained as follows.  Since the area made available to the flow upon just rising above the midplane is roughly constant, the continuity equation implies $\dot{m} \approx \rho v$.  A decrease in velocity will thus be accompanied by a slight increase in density (and hence temperature), in general.  The magnitude of this effect is dependent on the initial velocity.  For the $i = 60^\circ$ solutions, $v_o$ is only slightly below $c_o$ for the tail solution, so only a marginal increase in velocity is needed to reach the sonic point.  The increase in temperature then acts to prevent the flow from immediately becoming sonic.  Indeed, the $i = 30^\circ$ tail solution has a higher initial velocity than the $i=60^\circ$ normal solution, but the latter wind has a closer sonic point because the former wind undergoes more adiabatic heating for $\x \la0.2$.

It is natural to suppose that one of the degenerate solutions is unstable.  However, this might be difficult to uncover analytically because each solution, being transonic, has an associated regularity condition.  The regularity condition acts as a boundary condition to aid the stability analysis and its mere presence may indicate stability (Velli 2001).  Therefore, time dependent simulation may be a more appropriate tool for assessing the stability of these solutions.   

%===============================================================================
% DISCUSSION thru CONCLUSIONS
%===============================================================================

\section{Discussion}\label{discussion}
As it stands, we have solved and graphically analyzed the Eulerian equations for polytropic winds undergoing Keplerian rotation and traversing the geometry of \fig{modelgeometry2}.  To fully uncover the new aspects of our solutions, we must revisit the spherically symmetric Parker problem, as well as the `rotating' Parker problem (i.e. a Parker wind following trajectories that conserve specific angular momentum).  Analytical considerations of the behavior of Parker winds at higher $\gamma$ constitute the basis of our discussion.  Readers more interested in the observational implications of our results are referred to our summary, as this section is geared toward investigators interested in the mathematical properties of Parker winds.   

\subsection{The Enthalpy Deficit Regime} \label{enthalpydeficit}

\setlength{\unitlength}{1.0cm}
 \begin{figure*} 
\begin{picture}(6,8)(-6,0)
\put(4,-2){\includegraphics{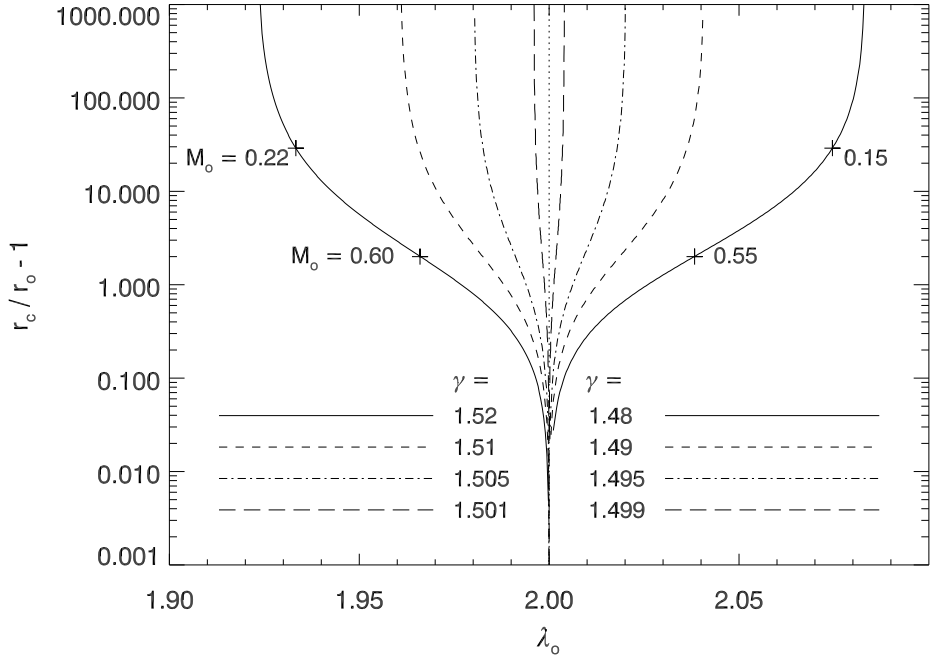}}
\end{picture}
\vspace{1 cm}
\caption{Location of critical (sonic) points for the spherically symmetric Parker problem as a function of the HEP in the neighborhood of $\gamma = 3/2$.   The vertical, dotted line at $\lp=2$ separates the enthalpy surplus and enthalpy deficit regimes.  Critical points curves in the latter regime (those with $\lp < 2$) have a negative slope (so that higher temperatures lead to more distant sonic points) and correspond to \emph{decelerating} transonic solutions, implying that $v_o > v_\infty$.}
\label{nearg1.5}
\end{figure*}

Depending on whether the sum of the enthalpy and the effective potential energy terms at the boundary is positive or negative, we define the flow as having an enthalpy surplus or deficit, respectively.  
By our dimensionless Bernoulli function, $e_c = y + (\lc/\lp)U_{eff} + h/c_s(\x_c)^2$, if the sum of the second and third terms is negative at $\x=0$, the kinetic energy $y$ must decrease as the sum becomes less negative at $\x>0$ to keep $e_c$ constant.  The flow can still be transonic because the sound speed (eventually) decreases faster than does the velocity.

Since $h(\x=0)/c_s(\x_c)^2 = s_o/(\gamma -1)$ and $U_{eff,o} = -\lp + \zeta^2/2$, where $s_o=\lc/\lp$, the defining condition for enthalpy deficit flow places the following requirement on the HEP:
\beq \lp > \f{1}{\gamma-1} + \f{\zeta^2}{2} . \label{edefregime}\seq
For Keplerian velocities $(\zeta^2 = \lp)$, this condition is simply $\lp > 2/(\gamma-1)$.
We see that this regime is encountered only for $\gamma$ sufficiently larger than 1.   From \fig{dwcritpts}, the Converging model enters this regime for $\gamma \geq 4/3$, while the CIA model only enters it for $\gamma \approx 5/3$.  

As discussed by Holzer \& Axford (1970), transonic winds require a positive Bernoulli constant, i.e. $e_c = B_o/c(\x_c)^2 > 0$.   The Bernoulli function thereby permits an alternative definition of the enthalpy deficit regime, namely $y_o > e_c$.  Equivalently, since $v_\infty/v_{esc} = \sqrt{e_c/\lc}$ (see \eqn{vterminal}), we must have that
\beq v_o > v_\infty .\seq
In terms of the initial Mach number, this lower bound reads $ \mathcal{M}_o > \sqrt{2e_c(\lp/\lc)}$.  In the classic Parker problem, the transonic bulk velocity profile is a monotonically decreasing function of $r$ for $\gamma > 3/2$ (see e.g., Lamers \& Cassinelli, 1999).  It follows that this class of solutions has $v_o > v_\infty$ and therefore lies in the enthalpy deficit regime.

The mathematical implication of there being a nonzero lower limit placed on $v_o$ or $\mathcal{M}_o$ naturally leads to the conclusion reached by Parker (1960) that viable solar wind solutions satisfy $1<\gamma<3/2$.  Only this class of solutions can have vanishingly small initial velocities, criteria that Parker imposed as a boundary condition.  For the same reason, the class of $3/2 < \gamma < 5/3$ solutions were overlooked in followup treatments, e.g. that of Carovillano \& King (1965).  Adhering to the physical assumption that the gas is launched from highly subsonic speeds automatically excludes the enthalpy deficit regime.  

Even if phenomenologically motivated, discounting the enthalpy deficit regime prohibits insight into the full nature of the problem.  Namely, we benefitted from the realization that Parker winds undergo a regime change because the flow must tap into its own kinetic energy to become transonic.\footnote{In stark contrast, spherically symmetric (Bondi) accretion flow is safely in the enthalpy surplus regime for all $\gamma$ since both the potential and velocity vanish at the boundary, taken to be infinity.}  It is informative to view the $\gamma$-dependence of the classic Parker problem from an enthalpy standpoint.  

\subsection{The Role of Enthalpy: Spherically Symmetric Parker Winds}\label{classicparker}
This much has been emphasized in textbook treatments of the classic Parker problem: $\gamma=3/2$ is the one value for which all of the enthalpy is used up to lift the gas out of the potential well, with none left over to supply kinetic energy (Lamers \& Cassinelli 1999).  The complete story is told by the dimensionless Bernoulli function, manipulated to read
\beq \f{\lc}{\lp} = \f{e_c}{\bigb{\f{1}{\gamma-1} - \lp} + \mathcal{M}_o^2/2} .\label{lclopar}\seq
We can look upon \eqn{lclopar} as either the temperature ratio $T_o/T_c$, or as representative of the critical point distance, as $r_c/r_o = \lc/2$.  Comparing the bracketed term in the denominator of \eqn{lclopar} with the allowed HEP ranges (given in \tbl{1}) allows us to infer the behavior of the critical point as $\gamma$ increases.  For transonic flows with a high energy input, meaning that $\gamma$ is close to 1, there is always a significant excess of enthalpy beyond that used to combat gravity that can contribute to increasing the kinetic energy of the gas.  Regardless of the HEP, $\mathcal{M}_o$ is small because the high energy input permits $v_o << c_o$.  The bracketed term is made smallest for large $\lp$, i.e., for lower coronal temperatures, so we have the intuitive notion that the smaller the coronal temperature, the more distant the sonic point and the smaller the initial Mach number.  

As $\gamma \rightarrow 3/2$ from below, progressively smaller HEP are required to compensate for the lower energy input into the wind as it expands.  This is shown by curves on the right half of \fig{nearg1.5}.  There remains a small enthalpy excess, so higher temperatures still lead to higher initial Mach numbers and less distant sonic points.  We show representative initial Mach numbers at $r_c/r_o = 3$ and $r_c/r_o = 30$ for $\gamma = 1.48$ to illustrate that  $\mathcal{M}_o$ can be almost as small as when $\gamma$ is much less than $3/2$, but only if the sonic point is very far away.  (For comparison, for $\gamma=1.1$, $r_c/r_o = 3$ corresponds to $\lp = 4.145$ and has $\mathcal{M}_o = 0.185$, while $r_c/r_o = 30$ corresponds to $\lp = 8.76$ and has $\mathcal{M}_o = 1.5\times 10^{-6}$).  Realistic solutions that traverse the sonic point within a few $r_o$ therefore require relatively high initial Mach numbers compared to those when $\gamma$ is small.  

For the special case $\gamma = 3/2$, the bulk velocity is a constant at all radii.  We show in \apx{newcritpts} that for this particular $\gamma$, the HEP is confined to the single value $\lp=2$, and the bulk velocity everywhere equals the initial sound speed, i.e. $\mathcal{M}_o = 1$ and therefore $r_c = r_o$.  The flow behavior for the class of accelerating solutions $1<\gamma<3/2$ can therefore be summarized as transitioning from an extreme surplus of enthalpy near $\gamma = 1$ that permits $\mathcal{M}_o \approx 0$ at low temperatures, to a scarcity of enthalpy near $\gamma = 3/2$ that, in order to get even barely accelerating transonic solutions, requires high temperatures and therefore high initial Mach numbers and close sonic points.

We then reach the high temperature regime ($>4\times 10^6K$ for solar parameters) of decelerating flow.  This class of critical point solutions with $3/2<\gamma<5/3$ was found by Dahlberg (1964) using a highly implicit formulation.  Parker, responding to this finding, stated ``it would be interesting to work out what conditions the solutions would fit to at the base of the corona where they start'' (Parker 1965).  Surprisingly, this never appears to have been done!  

We duly noted that the high temperature regime is characterized by an enthalpy deficit.  As shown by the curves to the left of $\lp = 2$ in \fig{nearg1.5}, critical point distances vary oppositely with HEP than for $\gamma < 3/2$; as the temperature decreases for a given $\gamma$, so does the sonic point distance, while the initial Mach number increases.  These features are easily explained.  Loosely speaking, since the flow is slowing down rather than speeding up, the gas can become sonic sooner only if the sound speed can quickly drop below the local magnitude of the bulk velocity, a scenario that is expedited if $c_o$ is smaller ($\lp$ larger) to begin with.  More rigorously, the bracketed term in \eqn{lclopar} is negative, so lowering the temperature (increasing $\lp$) forces the initial Mach number to be higher, as only a high launching velocity can compensate for the enthalpy deficit.  As a secondary effect, the temperature is so high that lowering it somewhat contributes to increasing $\mathcal{M}_o$.  Smaller sonic point distances follow.  We do note, however, by our representative values of $\mathcal{M}_o$  at $r_c/r_o = 3$ and $r_c/r_o = 30$ for $\gamma = 1.48$, that the initial Mach number in this regime does not greatly exceed $\mathcal{M}_o$ for $\gamma < 3/2$.  Indeed, $\mathcal{M}_o$ can be much less than 1, but only for very distant sonic points.  

\subsection{The Appearance of Velocity Minimums: Parker Winds with Rotation} \label{rotation}

\setlength{\unitlength}{1.0cm}
 \begin{figure*} 
\begin{picture}(6,8)(-6,0)
\put(4,-2){\includegraphics{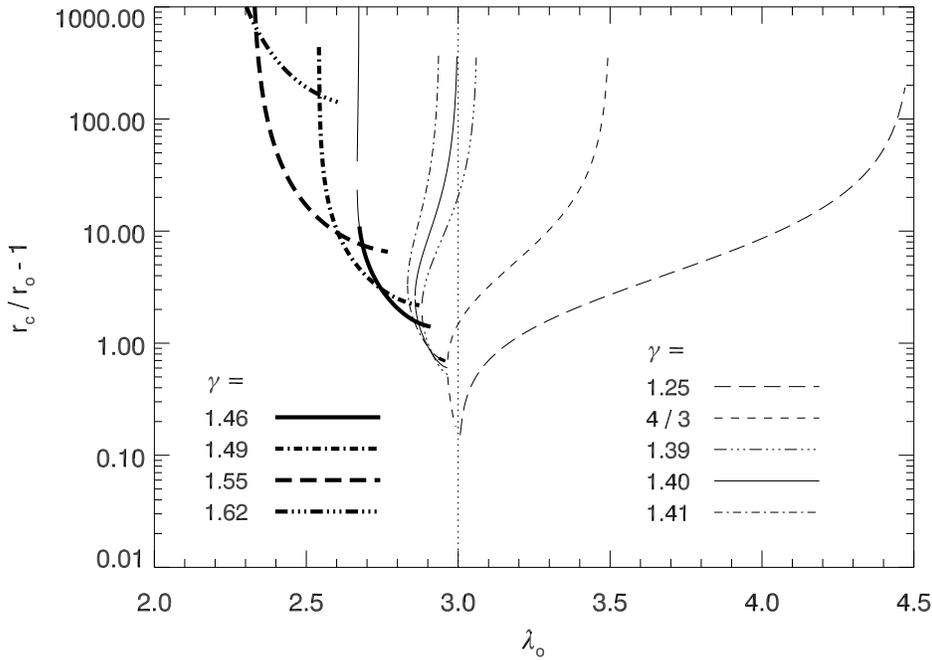}}
\end{picture}
\vspace{1 cm}
\caption{Location of critical (sonic) points vs. HEP for the Parker problem with rotation, where the (rigid) rotation rate is $\zeta = 1.0$.  This plot is to be compared with \fig{nearg1.5} which has $\gamma_t = 1.5$ and $\zeta=0$.   Here, $\gamma_t = 1.4$ and the enthalpy deficit regime sets in for HEP values to the left of the vertical dotted line at $\lp = 2+\zeta^2=3$, which divides transonic solutions with bulk velocity minimums from those without.  Only the bolded sonic points have transonic solutions in the enthalpy deficit regime with $v_o > v_\infty$. }
\label{zeta1vsg}
\end{figure*}

We can analytically investigate the effects of including angular momentum by considering the `rotating Parker problem', in which the streamlines are radial and spherically divergent but rotate rigidly with velocity $u_\phi = \zeta c_o$.  Since stars differentially rotate, this solution is valid near the equatorial plane only.  In a disc wind context, this solution can be viewed as a wind emanating from the edge of a flared, rigidly rotating disc; the differentially rotating, `Keplerian Parker wind' solution used by Adams et al. (2004) and Gorti \& Hollenbach (2009) corresponds to the special case $\zeta = \sqrt{\lp}$.  Results for the spherically symmetric Parker problem are recovered by setting $\zeta=0$.  

We begin by finding the relationship between $\rh_c$ and $\lc$, where we are calling $\rh = r/r_g = (r/r_o)/\lp$.  The effective gravitational force is $g=[1-(\zeta^2/\lp^2)/\rh]/\rh^2$ (see \eqn{geff}), so we have by \eqn{criteqn} that
\beq \f{\lc}{\lp}=\f{2\rh_c}{1-(\zeta/\lp)^2/\rh_c} \label{lcrcParker}. \seq
This equation is quadratic in $\rh_c$ with solution,
\beq \f{r_c}{r_o} = \f{\lc}{4}\bigb{1 \pm \sqrt{1-8\f{(\zeta/\lp)^2}{\lc/\lp}}} .\label{rcoverro}\seq
Only the positive root is satisfied by the location of the critical point.  Is there any meaning to the negative root?  In the isothermal ($\lc=\lp$) case, in which the positive root of \eqn{rcoverro} directly yields the location of the critical point, the negative root gives the radius where the bulk velocity reaches its minimum value:\footnote{We are only concerned with the locations of local minimums, but \eqn{rminoverro} yields the location of local maximums also.  Indeed, in the limit $\zeta\rightarrow 0$, $r_{min}=0$ is a velocity minimum for $\gamma < 3/2$ and a velocity maximum for $\gamma > 3/2$.  Note that purely decelerating flows, which are mathematically equivalent to having bulk velocity minimums at infinity (or maximums at $r_c/r_o < 1$), do not occur for sufficiently high rotation rates.}
\beq \f{r_{min}}{r_o} =  \f{\lp}{4}\bigb{1 - \sqrt{1-8\bigp{\f{\zeta}{\lp}}^2}} .\label{rminoverro}\seq
This occurrence can be accounted for mathematically by recalling the derivation leading up to \eqn{criteqn}.  As we discussed in \S\ref{singnreg}, the critical point must satisfy both the singularity and regularity conditions.  The latter defines $y'$ when $y=1/2$.  The former simply picks out all points for which the equation of motion does not depend on the acceleration, a condition that is also satisfied if the acceleration is 0, i.e. at points where the velocity is a local maximum or minimum.  In other words, while the location of the critical point must satisfy \eqn{lcrcParker}, there exist rotation rates when $\lc = \lp$ in which points where $y'=0$, not critical points where $y=1/2$ (and $y' = 0/0$ by \eqn{F0}), also obey this equation.

The equation governing the location of bulk velocity minimums in the polytropic case has the same form as \eqn{rminoverro}, but involves $s_{min} \equiv s(\x=\x_{min})$:
\beq \f{r_{min}}{r_o} = \f{\lc}{4 s_{min}}\bigb{1 - \sqrt{1-8\f{(\zeta/\lp)^2}{\lc/\lp}s_{min}}} .\label{rminpoly}\seq
In general, this location can only be solved for numerically, as $s_{min}$ is a function of both $w(\x_{min})$ and $A(\x_{min})$.  However, \eqn{rminoverro} still approximates this location when $r_{min} \la 2 r_o$ if the flow is close to being isothermal, say $\gamma \la 1.05$.  In that circumstance, $s_{min} \approx s_o = \lc/\lp$, and \eqn{rminpoly} reduces to \eqn{rminoverro}. 

Analyzing \eqn{rminpoly}, velocity minimums will not arise unless $r_{min}/r_o > 1$, implying the bound
\beq s_{min} > \f{\lc}{2}\bigp{1-\f{\zeta^2}{2}} .\seq
As it is, this inequality is not very insightful since $\lc$ is unknown, but since the flow will adiabatically cool, we must also have $s_{min} < s_o = \lc/\lp$.  Combining these two inequalities, we can conclude that for a given rotation rate, velocity minimums will not be present unless
\beq \lp < 2 + \zeta^2 .\label{lpcrit}\seq
This criteria follows more directly from the equation of motion, as it is the condition for the flow to decelerate off the midplane, i.e. for $y'(\x=0)<0$.  Hence, inequality (\ref{lpcrit}) is the statement that flows undergoing deceleration must be of very high temperature!  It certainly contradicts our intuitive notion that higher temperatures give rise to winds with steeper positive velocity gradients.  Evidently, the enthalpy deficit regime must have $\lp < 2 + \zeta^2$, and the appearance of velocity minimums is to the rotating Parker problem what purely decelerating flow is to the spherically symmetric Parker problem.  Recall that solutions in the latter problem with $\gamma > 3/2$ still have monotonically increasing Mach number profiles; never does the bulk velocity decrease faster than the sound speed, which would cause $\mathcal{M}$ to decrease.  An effect of rotation is to allow this happen.  Of course, the locations of Mach number minimums and bulk velocity minimums do not coincide.  It will be seen below that Mach number minimums imply accompanying bulk velocity minimums, but the converse is not true.

Notice that for Keplerian rotation ($\zeta^2 = \lp$), the inequality (\ref{lpcrit}) is always satisfied, explaining why the bulk velocity profiles of our disc wind solutions possess minimums at all $\gamma$.  
It remains to explain the behavior of the tails of the critical point curves.  For this, we must carefully study the enthalpy deficit regime of the rotating Parker problem.              

\setlength{\unitlength}{1.0cm}
 \begin{figure*} 
\begin{picture}(6,8)(-6,0)
\put(4,-2){\includegraphics{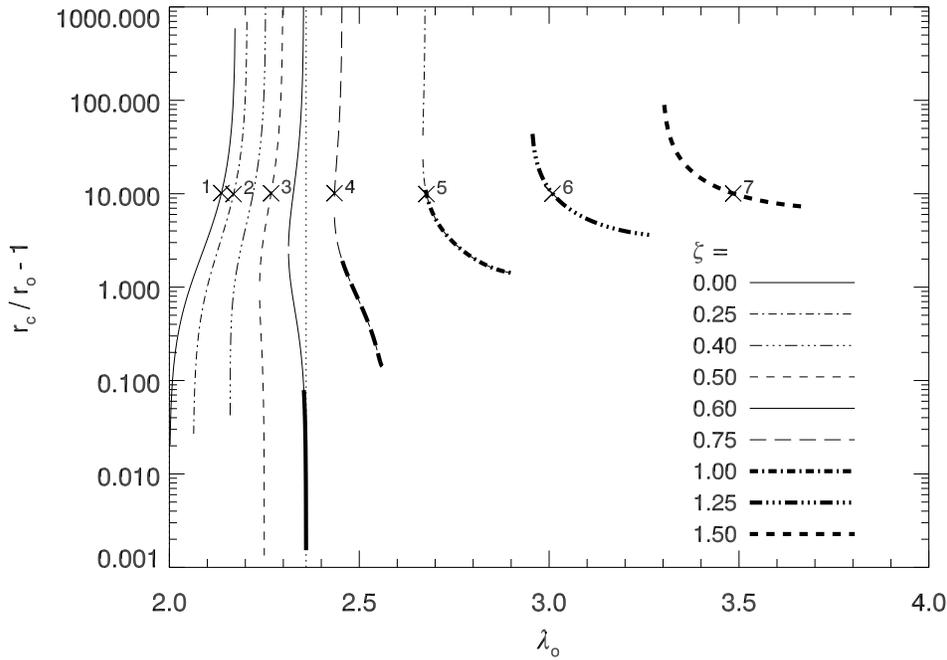}}
\end{picture}
\vspace{1 cm}
\caption{Location of critical (sonic) points vs. HEP for $\gamma = 1.46$ for various rotation rates $\zeta = u_\phi/c_o$ ranging from 0 up to 1.5 for the Parker problem with equatorial rotation.  The effect of rotation is to shift the enthalpy deficit flow regime to smaller $\gamma$, so that this regime (bold curves) is entered by holding $\gamma$ fixed and increasing $\zeta$.  We placed a vertical line at $\lp = 2.36$ to illustrate the tendency toward the upper bound $2+\zeta^2$.  The numbered crosses have the properties tabulated in \tbl{4} and correspond to the transonic solutions plotted in \fig{transonics}.  
}
\label{g1.46plot}
\end{figure*}

\begin{table*}
\footnotesize
\begin{center}
\caption{Properties of the $\gamma=1.46$ transonic solutions plotted in \fig{transonics}:}
\begin{tabular}{c l l l c c c c c c c}  \\ \hline
\hline 
Label & $\zeta$ & $\gamma_t$ & $\lp$  & $(\lp)_{crit}^*$ & $r_c/r_o$  & $\lc$ & $e_c$ & $\mathcal{M}_o$ &$v_o/v_{esc}$ & $v_\infty/v_{esc}$\\
\hline
 (1) & 0.  & 1.5  & 2.137 & 2.1739& 11.1673 & 22.3347 & 0.6739 & 0.2348 & 0.1136 & 0.1737  \\
 (2) &0.25  & 1.4923 & 2.169 & 2.2052 &10.9180 & 21.8939 & 0.6713 & 0.2463 & 0.1183 & 0.1751 \\
 (3) & 0.5    & 1.4706 & 2.268 & 2.2989 & 11.1108 & 22.4442 & 0.6639 & 0.2690 & 0.1263 & 0.1720  \\
 (4) &$0.75^\dagger$   & 1.4384  & 2.435  & 2.4551 & 11.1302 & 22.7323 & 0.6527 & 0.3154 & 0.1429 & 0.1694 \\  
 (5) &1.0  & 1.4 & 2.677 & 2.6739  & 10.9287 & 22.6310 & 0.6385 & 0.3965 & 0.1714 & 0.1680  \\
 (6) & 1.25 & 1.3596 & 3.009  & 2.9552  &  11.0326 & 23.1550 & 0.6245 & 0.5196 & 0.2118 & 0.1642  \\
 (7) & 1.5  & 1.32 & 3.485 & 3.2989  & 11.0584 & 23.4881 & 0.6119 & 0.7441 & 0.2819 & 0.1614 \\
\hline \hline 
\end{tabular}
 \end{center}
 \begin{center}
 $^*\,(\lp)_{crit} = 1/(\gamma-1) + \zeta^2/2$ \\
 $^\dagger$For $\lp = 2.435$, there are two wind roots; the second has $r_c/r_o = 6.2391$, $\mathcal{M}_o = 0.4455$, \& $\lc=12.958$   \\
\end{center}
\normalsize
\label{zetatable}
\end{table*}

\setlength{\unitlength}{1.0cm}
 \begin{figure*} 
\begin{picture}(6,10.5)(-6,0)
\put(3.5,2.95){\includegraphics{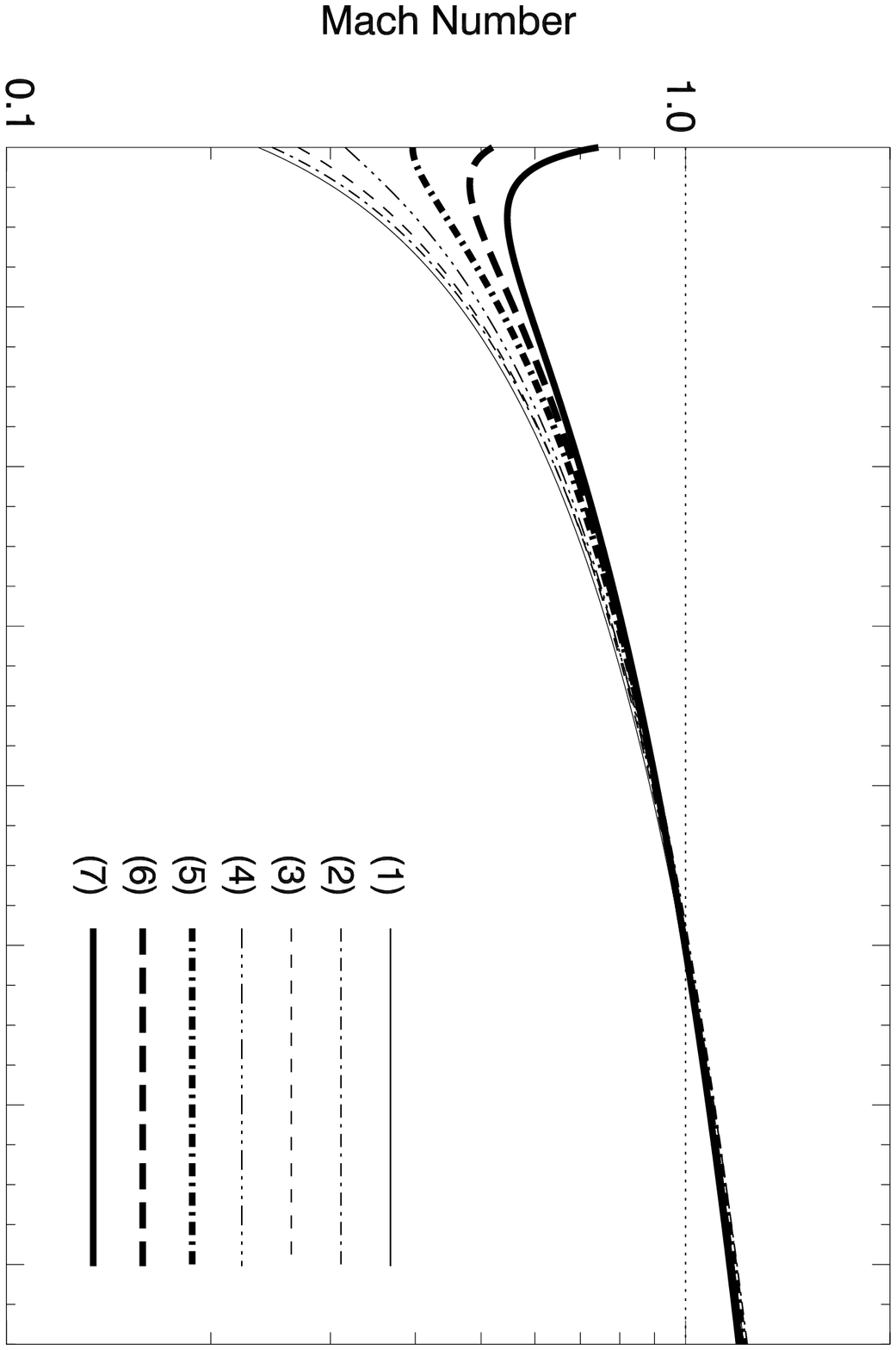}}
\put(3.5,-2.5){\includegraphics{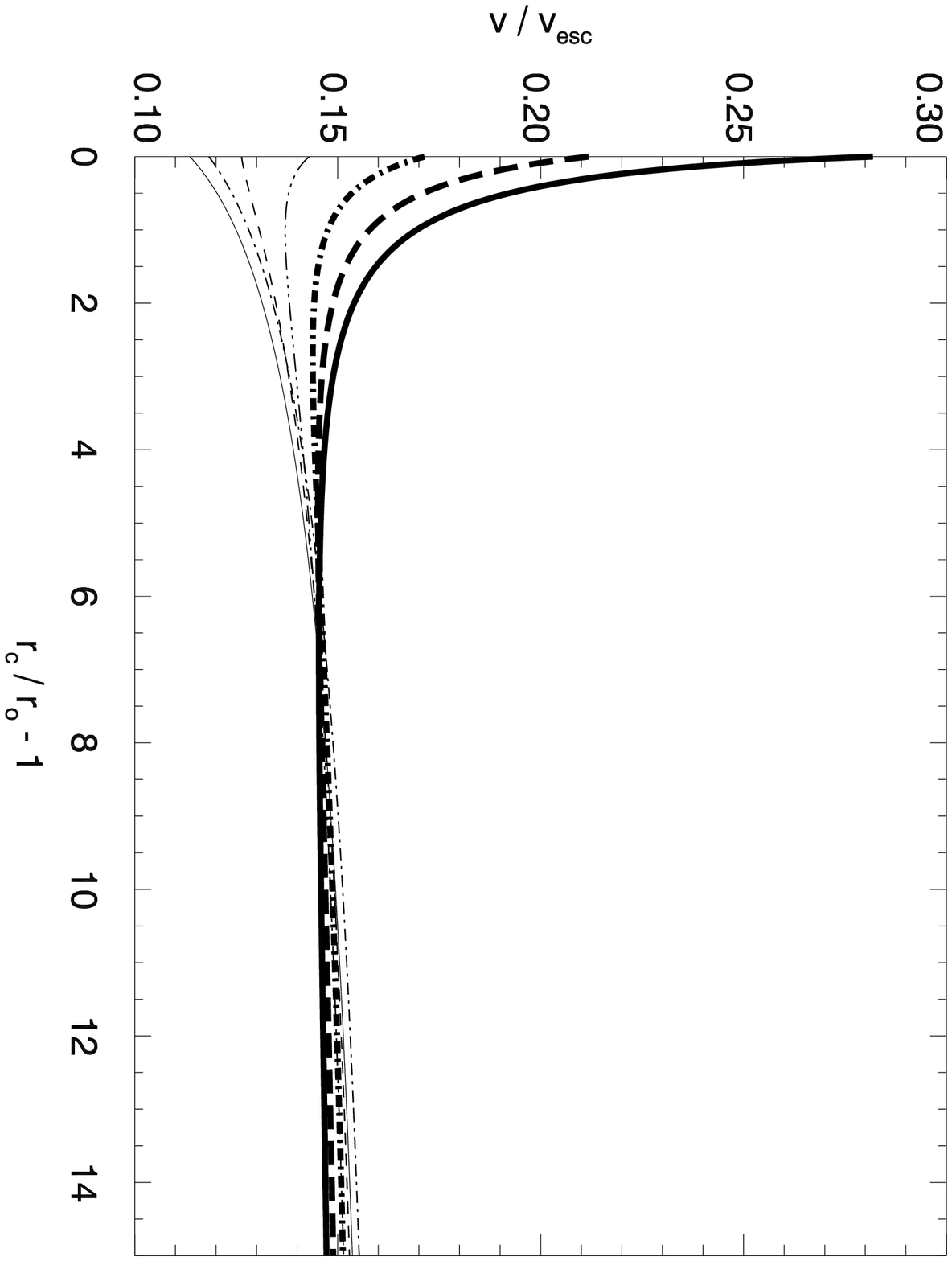}}
\end{picture}
\vspace{1.5cm}
\caption{Transonic Mach number (top) and bulk velocity (bottom) profiles for the rotating Parker problem with $\gamma = 1.46$.  The numbered solutions have the properties listed in \tbl{4}; each solution has a sonic point near $r_c/r_o = 11$.  Solutions (5)-(7), bolded, are in the enthalpy deficit regime with $\lp >1/(\gamma-1) + \zeta^2/2$ and $v_o > v_\infty$.  In contrast to the spherically symmetric Parker problem, bulk velocity profiles are not monotonically decreasing in this regime.  The bolded curves also have Mach number minimums, although this is due more to the increased rotation rate, i.e. enthalpy surplus solutions can also display Mach number minimums if $\zeta \ga 1$.  
}
 \label{transonics}
\end{figure*}

\subsection{The Enthalpy Deficit Regime with Rotation}\label{edefrotation}

The value $2+\zeta^2$ is the critical HEP below which the bulk velocity must initially decelerate before becoming transonic.  Each rotation rate must correspondingly lead to a unique polytropic index, $\gamma_t$ say, that determines the onset of enthalpy deficit flow.  By the inequality \ref{edefregime},
\beq \gamma_t = 1 + \f{1}{2 + \zeta^2/2} .\seq
Necessary (but not sufficient) criteria to reach the enthalpy deficit regime is for \emph{both} $\lp < 2+\zeta^2$ and $\gamma > \gamma_t$.  Sufficient criteria is, by definition, that $\lp > 1/(\gamma-1) + \zeta^2/2$. The inclusion of rotation lowers $\gamma_t$ below $3/2$, which may seem counterintuitive but is a consequence of a tradeoff between the enthalpy and the effective potential.  Consider two winds with the same total energy, but only one with a boost from the centrifugal force.  The wind with rotational energy must be launched with a smaller temperature, and so it must in turn have a smaller enthalpy at the base ($h \propto T$).  However, a smaller temperature tends to increase the magnitude of the effective potential at the base (because $U_{eff,o} = -\lp + \zeta^2/2$ and $\zeta^2 \leq \lp$).  Since the requirement for the enthalpy deficit regime can be restated as $s/(\gamma - 1) < |U_{eff,o}|$, a larger $|U_{eff,o}|$ allows for a smaller $\gamma$.   

Another effect of rotation is to smooth the transition from `normal' critical point behavior, that with $r_c/r_o$ increasing with HEP, to enthalpy deficit behavior.  This situation is shown in \fig{zeta1vsg}, which is the rotational analogue to \fig{nearg1.5}.  In both figures, $\lp = 2+\zeta^2$ marks the critical HEP below which $r_c/r_o$ begins to decrease as $\lp$ increases.  
An obvious difference is that in \fig{zeta1vsg}, two wind solutions arise for the same HEP to left of the vertical dotted line, reminiscent of our disc wind results.  Only the (parts of) curves that are bolded are in the enthalpy deficit regime, having $\lp > 1/(\gamma-1) + \zeta^2/2$.  (The tail of the $\gamma = 1.41$ curve is the first to meet this criteria.)  As $\gamma$ is futher increased, the tail grows in length and the normal critical points begin to disappear altogether.  

The HEP bounds differ according to this behavior.  If $\gamma <\gamma_t$ and there is only one transonic solution per HEP (i.e. if the curve has no tail), then
\beq 2 + \zeta^2 < \lp<\f{1}{\gamma-1} + \f{\zeta^2}{2} .\seq
Whenever there are two critical points per HEP, the normal critical points have
\beq \lambda_{o,min}(\gamma,\zeta) < \lp<\f{1}{\gamma-1} + \f{\zeta^2}{2} .\seq
The minimum HEP attained, $\lambda_{o,min}(\gamma,\zeta)$, appears to be a priori unknown and is a function of both $\gamma$ and $\zeta$.  The points lying on the tail, meanwhile, have
\beq \lambda_{o,min}(\gamma,\zeta) < \lp< 2 + \zeta^2 .\seq 
Since $\gamma_t$ is the polytropic index at which $1/(\gamma-1) + \zeta^2/2=2+\zeta^2$, it marks where the upper bounds switch places.
Only once $\gamma$ is high enough above $\gamma_t$ such that there is again only one critical point per HEP do we encounter the fully enthalpy deficit regime.  In that case, the tail becomes the entire critical point curve and the HEP bound is
\beq \f{1}{\gamma-1} + \f{\zeta^2}{2} < \lp< 2 + \zeta^2 \label{upperbounds}.\seq
Note that we recover the bounds reported in \tbl{1} when $\zeta = 0$.  

Another look at the transition from enthalpy surplus to deficit flow given in \fig{g1.46plot}.  Rather than changing $\gamma$ for a fixed rotation rate, we set $\gamma=1.46$ and vary the rotation rate from 0 from to $\zeta = 1.5$.  In this way, the enthalpy deficit regime is gradually reached as $\gamma_t$ drops from 1.5 for $\zeta=0$ to 1.32 for $\zeta = 1.5$.  Notice that the normal (enthalpy surplus) critical point curves, the first of which begins at $r_c/r_o \approx 1.02$ for $\zeta = 0$, steadily shift upward to begin and end at higher sonic point distances as the rotation rate is increased.  Slightly past $\zeta = 1$, however, the HEP range becomes maximally confined by $\lambda_{o,min}(\gamma,\zeta) < \lp<  2+\zeta^2$, and only enthalpy deficit roots are allowed.  The latter make an appearance on the tail of the $\zeta = 0.6$ curve, the first curve to have $\lp < 2 + \zeta^2 = 2.36$ and $\gamma > \gamma_t =  1.4587$.       

\fig{g1.46plot} reveals that, at a fixed energy input (i.e. constant $\gamma$) near the enthalpy deficit regime, increasing the rotation rate steadily increases the minimum HEP for which critical point solutions exist and pushes the lowest sonic point distance to higher values.  That is to say, the centrifugal force permits transonic solutions to arise at lower temperatures overall.  The flow remains subsonic out to progressively larger radii because a balance must be struck between keeping the flow subsonic at these (still high) temperatures under low effective gravity and simultaneously supplying enough energy to launch a transonic wind when there is little energy injected into the flow at these high $\gamma$. 

We have selected critical points at $r_c/r_o \approx 11$ in \fig{g1.46plot} to illustrate the effect that rotation has on the Mach number profiles.  These critical point solutions have the properties listed in \tbl{4} and are plotted in \fig{transonics}.  Curves with higher rotation rates always have larger initial Mach numbers, as expected.  Bulk velocity minimums occur whenever the HEP is less than $2+\zeta^2$, met by solutions (4) and higher, while we see that the Mach number profiles have minimums only for the three solutions in the enthalpy deficit regime, namely (5), (6) and (7).  However, having $v_o > v_\infty$ is not a necessary condition for there to be a Mach number minimum, only a sufficient one.  Indeed, all critical points on the tail curves exhibit Mach number minimums at the rotation rates that we sampled.  At even higher rotation rates, Mach number minimums occur for points on the normal critical point curves.  The appearance of Mach number minimums for our disc wind solutions is therefore solely an effect of the high Keplerian rotation rate.  

\section{Summary \& Conclusions}\label{conclusions}
We have studied generalized solutions of the classic Parker problem by treating spherical and cylindrical geometries in a unified fashion.  Parker-like disc winds differ from Parker winds in that (i) varying the HEP for Parker winds samples different coronal temperatures for a given $M_*$, whereas varying the HEP for disc winds samples different temperatures as well as distances along the equatorial plane; and (ii) the purely decelerating wind regime for Parker winds with $\gamma > 3/2$ is replaced by an initially decelerating wind that reaches a minimum speed and then proceeds to accelerate for $l>r_o$.  We showed that the equivalent nozzle function can be used as a means to gauge whether or not the velocity is monotonic without actually finding the transonic solutions.  Meliani et al. (2004) also employed equivalent nozzle functions in their relativistic generalization of the polytropic Parker problem using the Schwarzschild metric.  There, winds were also found to accelerate for $\gamma>3/2$, but the velocity profiles were still monotonic.  We showed velocity minima to be an effect of adding rotation. 

Our discussion of the spherically symmetric and rotating Parker wind solutions showed that significant deceleration is associated with a flow regime characterized as having an enthalpy deficit.  It is not inconceivable that this type of outflow can exist in an astrophysical setting, so it would be interesting to determine the spectral signatures of a decelerating wind region.  We tied the enthalpy deficit regime (i.e. the parameter space giving solutions with $v_\infty < v_o$) to the appearance of degenerate transonic wind solutions, and we further pointed out that the critical point behavior of the second set of transonic solutions is remarkably similar to that reported by Cur\'e (2004) in his study of \textit{isothermal} line-driven stellar wind equations with rotation.  Cur\'e (2004) classified his new solutions as `slow', since they obtain significantly smaller terminal velocities.  Might these slow solutions be a different guise of the enthalpy deficit regime?  

Our main objective was to investigate the dynamical properties of two axially symmetric, thermally driven disc wind models, one with significant adjacent streamline divergence (the Converging model) and another with a complete lack thereof (the CIA model).  We emphasize that detailed hydrodynamical simulations show that by taking into account the interactions of neighboring flow tubes, a self-similar streamline geometry emerges (recall \fig{stefans}).  We have neglected to mention elsewhere that CIA-like streamlines have also been found analytically from similarity solutions of idealized models for galactic superwinds.  Both the self-similar solutions of Bardeen \& Berger (1978), which took into account a gravitational potential, and those of Zirakashvili \& V\"olk (2006), which did not involve gravity, are examples.  

Since the CIA and Converging models only differ by their respective amounts of streamline divergence, we can attribute the differences in the properties of their solutions as being solely due to geometric effects.  Our results have implications for kinematic models that adopt a flow geometry similar to the Converging model for the purposes of computing synthetic spectra to compare with observations.  Namely, use of the Converging model will significantly overestimate the acceleration of the flow if the true wind configuration more closely resembles the CIA model.  The latter model, due to its smaller amount of streamline divergence, features a greatly extended acceleration zone, a more distant sonic surface, a shallower density and temperature falloff, and a smaller mass flux density than the Converging model for similar footprint conditions.  Conversely, for a given mass flux density at the wind base, the CIA model will predict a higher density, implying that synthetic line profiles will exhibit stronger line absorption.  Ultimately, larger error bars may need to be associated with the inferred mass-loss rate, as spherically diverging winds may tend to over-estimate it.

Solving the time-\textit{dependent} problem is likely to yield insights into the full domain of viable disc wind solutions.  Especially considering that we found degenerate transonic solutions, uncovering the effects of time-dependence is a worthwhile task, one that we plan to undertake in a future work.  It is likely that one of the degenerate solutions is unstable, and this could be verified using hydrodynamical simulations.  The alternative would be more exciting, however, as it is conceivable that the time-dependent solution can settle upon both solutions under various circumstances.  

%===============================================================================
% REFERENCES
%===============================================================================

\bigskip
\noindent\textbf{Acknowledgments}\\
We thank the referee for many helpful suggestions that improved the quality and presentation of this paper.
We acknowledge support provided by the Chandra award TM0-11010X issued by the Chandra X-ray Observatory Center, which is operated by the Smithsonian Astrophysical Observatory for and on behalf of NASA under contract NAS8-39073.  Funding for TRW was provided by the Nevada NASA Space Grant Consortium, through NASA grant NNX10AJ82H.  DP also acknowledges the UNLV sabbatical assistance and support from Program
number HST-AR-12150.01-A that was provided by NASA through a grant from the Space Telescope Science Institute, which is operated by the Association of Universities for Research in Astronomy, Incorporated, under NASA contract NAS5-26555.
TRW is indebted to DP for moral support and expert guidance, and he thanks Eugene Parker for being an inspirational figure in astrophysics.  

\nocite{*}
\bibliographystyle{mn2e}
%\bibliography{references}

%===============================================================================
% APPENDICES
%===============================================================================
\appendix
\section{Formulae for the Polytropic Solution} \label{formulae}
To facilitate the usage of our solutions for numerical testing purposes, we first sketch the solution procedure.  All quantities given in Appendix A follow algebraically from the Mach number, $\mathcal{M}=\sqrt{w}$, which can be determined numerically from the explicit solution, \eqn{explicit}, for $1<\gamma\leq 5/3$.  Since there are multiple values of $\x_c$ satisfying \eqn{critpointeqn}, a rootfinder should screen for valid wind solutions by checking which roots satisfy $\mathcal{M}_o < 1$ according to \eqn{wo}.  For every critical point $\x_c$, there is a unique value of $\lc$ given by \eqn{criteqn}.

\subsection{The Bulk Velocity, Sound Speed, Density, \& Internal Energy Profiles}
We use as a characteristic velocity, $v_{esc} = \sqrt{2GM_*/r_o}$, in terms of which the HEP is given by $\lp = v_{esc}^2/2c_o^2$.  For disc wind solutions, simply make the substitution  $v_{esc} \rightarrow \sqrt{2}V_{esc}$, as the appropriate escape velocity for a Keplerian disc is $V_{esc} = \sqrt{GM_*/r_o}$. 

The bulk velocity is obtained from the specific kinetic energy $y=\lc (v/v_{esc})^2/2$, which can be found by eliminating $s$ from \eqref{eigval} via $y =sw/2$, giving
\beq \f{v}{v_{esc}} = \sqrt{\f{1}{2\lc}} \bigp{\Lambda\f{\mathrm{A}_c}{\mathrm{A}}}^\igfacb \mathcal{M}^\f{2}{\gamma+1} .\label{ubulk}\seq
The sound speed is then simply $c_s/v_{esc} = (u/v_{esc})/\mathcal{M}$ by definition, or explicitly from \eqref{eigval},
\beq \f{c_s}{v_{esc}} = \sqrt{\f{1}{2\lc}} \bigp{\Lambda\f{\mathrm{A}_c}{\mathrm{A}}\f{1}{\mathcal{M}}}^\igfacb  .\label{csovervesc}\seq
The above equations reduce to identities in the isothermal $\gamma=1$ case in which $\lc=\lp$.  

The density follows straightforwardly from the polytropic EoS,
\beq \f{\rho}{\rho_o} = \bigp{\sqrt{2\lp}\f{c_s}{v_{esc}}}^{\f{2}{\gmone}} = \bigp{\f{\lp}{\lc}}^\f{1}{\gmone}\bigp{\Lambda\f{\mathrm{A}_c}{\mathrm{A}}\f{1}{\mathcal{M}}}^\f{2}{\gpone} ,\label{polydensity}\seq
where the second equality makes for an interesting comparison with the isothermal result, \eqn{rhoiso}.  Note that $\rho_c/\rho_o = (\lp/\lc)^{1/(\gamma-1)}$. 
The temperature profile is readily obtained from \eqn{csovervesc} but is most simply expressed in terms of the density profile as $T/T_o = (\rho/\rho_o)^{\gamma-1}$.

Finally, the internal energy density $E$ is found from $P=(\gmone)E$ combined with $c_s^2=\gamma P/\rho$:
\begin{align}
 \f{E}{\rho_o v_{esc}^2} &= \f{1}{\gamma(\gmone)}\f{\rho}{\rho_o}\bigp{\f{c_s}{v_{esc}}}^2 \\
 &= \f{1}{2\gamma(\gmone)}\bigp{\f{\lp}{\lc^\gamma}}^\f{1}{\gmone}\bigp{\Lambda\f{\mathrm{A}_c}{\mathrm{A}}\f{1}{\mathcal{M}}}^\f{2\gamma}{\gpone}.\label{Epoly}
 \end{align}
In terms of the footprint value $E_o \equiv E(\x=0)$, the internal energy density is simply
\beq \f{E}{E_o} = \bigp{\Lambda\f{\mathrm{A}_o}{\mathrm{A}}\f{\mathcal{M}_o}{\mathcal{M}}}^\f{2\gamma}{\gpone} .\seq
Using \eqn{csovervesc}, we find that at the critical point, $E_c/E_o = (\lp/\lc)^{\gamma/(\gamma-1)}$. 

\subsection{The Mass Loss Rate, Initial Velocity, \& Terminal Velocity}
From equations (\ref{CeqnD}) and (\ref{seqnD}) evaluated at the critical point, we find the critical mass flux density
\beq \dot{m}_c  = \rho_o c_o \f{\mathrm{A}_c}{A_o} \bigp{\f{\lp}{\lc}}^\f{\gpone}{2(\gmone)} .\label{mdot}\seq
Recalling that $A_o$ is a differential flow area, the total mass loss rate is found from $\dot{M} = \int \dot{m} A_o$, where the integral is taken over the wind region on the disc midplane ($A_o = 2\pi r_o dr_o \sin i$) or spherical boundary ($A_o = 2\pi r_o^2\sin \theta d\theta$).  Since $\dot{m} = \rho_o c_o \mathcal{M}_o$, the initial Mach number is $\mathcal{M}_o = (\mathrm{A}_c/\mathrm{A}_o)(\lp/\lc)^\f{\gpone}{2(\gmone)}$,
from which we get the initial velocity in escape speed units ($c_o = v_{esc}/\sqrt{2\lp}$):
\beq \f{v_o}{v_{esc}} = \sqrt{2\lp}\f{\mathrm{A}_c}{\mathrm{A}_o}\bigp{\f{\lp}{\lc}}^\f{\gpone}{2(\gmone)} \label{uoovervesc} .\seq
Equations (\ref{mdot}) and (\ref{uoovervesc}) both apply to $\gamma =1$ when casted in terms of the density using $\rho_c/\rho_o = (\lp/\lc)^{1/(\gamma-1)}$.
The terminal velocity is found by evaluating \eqn{BeqnD2} at infinity, where both the effective potential and pressure (and hence $s$) vanish, giving
\beq \f{v_\infty}{v_{esc}} = \sqrt{\f{e_c}{\lc}} .\label{vterminal}\seq

\section{The Bondi Problem}\label{Bondiproblem}
It is instructive to apply our dimensionless formulation to the classic Bondi problem, in which the boundary conditions are evaluated at infinity.  As mentioned at the beginning of \sec{Dimensionless}\hspace{4pt}, $r_g\rightarrow r_B = GM/c_\infty^2$ in that limit.  With $\rh=r/r_B$, the potential is simply $U=-1/\rh$ (and so $g=1/\rh^2$), and the critical point must obey \eqn{criteqn}, giving
\beq \rh_c=\f{1}{2}\f{\lc}{\lp} \label{lcrcbondi}. \seq
In light of \eqn{lcrcbondi}, the potential at the critical point can be written as $U_c = -2\lp/\lc$.  Inserting this into \eqn{ecpoly} gives
\beq e_c = \f{1}{2}\bigp{\f{5-3\gamma}{\gmone}}  .\label{egbondi2}\seq 
Equations \eqref{lcrcbondi} and \eqref{egbondi2} are the critical point conditions and are the same as those of the classic Parker problem. 

Determining an explicit expression for the location of the critical point requires knowing both the critical point constant, $e_c$, and the Bernoulli constant, $B_o$.  We can then eliminate $\lc/\lp$ from \eqn{lcrcbondi} by recalling that $\lc/\lp = e_c/(B_o/c_o^2)$.  Equivalently, $\lc/\lp$ can be found directly be evaluating the dimensionless Bernoulli function, \eqn{BeqnD2}, at infinity; we require $y_\infty=U_\infty=0$, and so
\beq \f{\lc}{\lp}= \f{5-3\gamma}{2}. \label{lcrcbondi2} \seq
Since $\lc/\lp = c_\infty^2/c_s(r_c)^2$, \eqn{lcrcbondi2} gives a pre-determined relationship between the sound speed at the critical point and the sound speed at the boundary, a situation unique to the Bondi problem.  It is a consequence of letting $w_\infty=0$ and is contrary to wind problems in which $w_o$ is nonzero and tied to the location of the critical point.  With \eqn{lcrcbondi2} substituted into \eqn{lcrcbondi}, we have $\rh_c=(5-3\gamma)/4$, which in physical units is the well known result
\beq r_c = \f{GM}{ c_{\infty}^2}\f{5-3\gamma}{4} .\label{rcbondi}\seq

The Bondi accretion rate is recovered from \eqn{mdot}, which in terms of the critical differential mass-loss rate reads
\beq \f{d\dot{M}_c}{\rho_\infty c_\infty}  =  \mathrm{A}_c \bigp{\f{\lp}{\lc}}^\f{\gpone}{2(\gmone)} .\label{mdotbondi}\seq
With $A_c =2\pi \rh_c^2\, r_B^2 \sin \theta d\theta$, we obtain after substituting in equations (\ref{lcrcbondi}) and (\ref{lcrcbondi2}),
\beq \f{d\dot{M}_c}{2 \pi r_B^2 \sin \theta d\theta \rho_\infty c_\infty}  = \f{1}{4}\bigp{\f{5-3\gamma}{2}}^{-\bigp{\f{5-3\gamma}{2(\gamma-1)}}}.  \seq
For the isothermal ($\gamma = 1$) case, $\rh_c = 1/2$ and so the Bondi rate is given by \eqn{gammaB} as $\Gamma_B= (2\pi r_B^2 \sin \theta d\theta/A_o)e^{3/2}/4$.  With $d\dot{M}_c = \rho_o c_o A_o \Gamma_B$, we see that there is no issue with calculating $\dot{M}$ despite $A_o \rightarrow \infty$ because there is no actual dependence on the area at the boundary.  In other words, we formally have $\Gamma_B = 0$ in the Bondi problem, as it is simply the initial Mach number of the flow (since $d\dot{M}_c = \rho_\infty c_\infty A_\infty \mathcal{M}_\infty$, which is still valid at infinity because the product $A_\infty \mathcal{M}_\infty$ is finite).

The isothermal Mach number profile is, from \eqn{Wfctsoln},
\beq \mathcal{M}(\rh) =\sqrt{ -W\bigb{-(\Lambda e^{3/2}/4)^2 \bigp{ \f{\exp[-1/\rh]}{\rh^2} }^2 } }, \label{Moisobondi} \seq
and the corresponding density distribution is
\beq \f{\rho(\rh)}{\rho_o} = \f{\Lambda e^{3/2}/4}{\rh^2\mathcal{M}(\rh)} .\label{rhoisobondi}\seq
The X-type solution topology for various values of $\Lambda$ can readily be explored by plotting equations \eqref{Moisobondi} and \eqref{rhoisobondi}.

\section{The Continuity Equation}\label{ctyeqn}
\vspace{-5pt}
Here we derive the continuity equation appropriate for any axially-symmetric streamline geometry consisting of straight streamlines in the $(x,z)$-plane.  See Fukue (1990) for the appropriate generalization when streamlines possess curvature.
In stellar wind equations with spherical symmetry, the continuity equation is usually stated as 
\beq \dot{M}=4\pi r^2 \rho(r)v(r) .\label{pvA}\seq
The task is to arrive at the appropriate differential area function $A(l)$ that yields the area between streamlines for the disc wind geometry of \fig{modelgeometry2}.  The continuity equation then takes the form
\beq d\dot{M}=\rho(l)v(l)A(l) .\seq

We begin with the steady state continuity equation in its coordinate-free, differential form, 
\beq \div \v{j} = 0, \seq
where $ \v{j}\equiv \rho \v{v}$.  Hence by the divergence theorem, 
\beq \int_V \div \v{j} \, dV=\oint_S \v{j}\cdot\hat{n}\,dS  ,\seq
the area occupied by streamlines pointing along $\v{j}$ that cross a surface $S$ with surface normal $\hat{n}$ is equivalently obtained from
\beq \oint_S \v{j}\cdot\hat{n}\,dS =0.\label{surfaceintegral} \seq
The spherically symmetric case ($\v{j} \cdot \hat{n} = j(r)$) is easily treated by considering the flow passing through two small solid angles $d\Omega_1$ and $d\Omega_2$ at two different radii $r_2 > r_1$.  The surface integral must vanish at both $r_1$ and $r_2$ separately, so
\beq j(r_1)r_1^2d\Omega_1 = j(r_2)r_2^2d\Omega_2. \seq
Since these radii are arbitrary, each side must equal a constant, $d\dot{M}$, so that at any radius,
\beq d\dot{M} = \rho(r)v(r)r^2d\Omega.\seq
To capture the entire flow area giving $\dot{M}$, we recover equation (\ref{pvA}) by integrating over all solid angles at constant $r$.  

Generalizing to the area contained between two wind cones at some height along the z-axis, we now work in cylindrical coordinates in which $\hat{n}=\hat{z}$, $\v{j}(l)\cdot \hat{z}=j(l)\cos (\pi/2 - i) = j(l)\sin i$, and $dS = xdxd\phi$.  Evaluating the surface integral in equation (\ref{surfaceintegral}) on a circular slice at some arbitrary height $z_1$ between two wind cones $x_2>x_1$ gives
\beq d\dot{M} = \int_{l=const}\v{j}(l)\cdot\hat{n}\,dS = j(l)\sin i \bigp{\pi x^2}\Big\lvert_{x_1}^{x_2}.\seq
From \fig{modelgeometry2}, we see that the integration limits are from $x_1=r_o+l\cos i$ to $x_2=x_1+dr_o+dr_i$, where the $dr_o$ step sweeps out the increase in area from moving further out along the disc, while $dr_i$ tracks the area swept out from streamline divergence alone.  The latter can be related to $di$ by projecting the arc length distance $l\,di$ onto the bold horizontal line:
\beq dr_i = -\f{ldi}{\sin i} .\seq
The negative sign accounts for the decrease in the angle $i$ farther out along the midplane, as we take $di$ to be positive.
After some algebra we arrive at
\beq A(l)= 2\pi dr_o(r_o + l\cos i)\sin i\bigb{1- \f{l(di/dr_o)}{\sin i}} .\label{streamlineA}\seq
This formula for the differential area traversed by the flow between two straight neighboring streamlines with an arbitrary amount of streamline divergence was obtained by Feldmeier \& Shlosman (1999) -- see their equation [19].\footnote{Note that this is only half the flow area needed to account for a biconical wind and hence to compute the actual mass loss rate.}  

\section{Origin of the `Tails' on the Critical Point Curves}\label{newcritpts}
In short, the tails on the critical point curves in \fig{dwcritpts} are due to the existence of the second root to the critical point equation, \eqn{critpteqn}.   
Insight into the nature of these roots can be gleaned from the spherically symmetric Parker problem, for which we can arrive at \eqn{critpteqn} by combining two separate relationships between the location of the critical point and the initial Mach number.  
The first is equivalent to equation [6] in Keppens \& Goedbloed (1999), whose quoted solutions we used to verify our numerical results.  It follows from the singularity condition, \eqn{criteqn}, taken together with \eqn{wo}, the combined polytropic/continuity relation.  Again calling $\rh = r/r_g$, the effective gravitational force is simply $g=1/\rh^2$, and $a=\rh^2$, as in the Bondi problem.  Subsituting $\lp/\lc = 1/2\rh_c$ into \eqn{wo} gives, since $\mathrm{A}_o= f ^2=\lp^{-2}$,
\beq \mathcal{M}_o = 2^{-\f{1}{2}\f{\gpone}{\gmone}}\lp^2 \rh_c^{-\f{5-3\gamma}{2(\gmone)}}.\label{rcMo1}\seq
A second relation between $\rh_c$ and $\mathcal{M}_o$ is found from \eqn{lclopar}:
\beq \rh_c = \f{(5-3\gamma)/4}{(\gamma-1)\bigb{\bigp{\f{1}{\gamma-1}- \lp} + \mathcal{M}_o^2/2 } }.\label{rcMo2}\seq 
As a first indication that these equations, when combined, possess multiple critical points, consider the special case $\gamma = 3/2$.  Equations \eqref{rcMo1} and \eqref{rcMo2} reveal that
\beq \rh_c = \f{1}{8}\bigp{\f{\lp}{2}+1}\bigb{\bigp{\f{\lp}{2}}^2 + 1}  \label{rc32}.\seq
Substituting this back into \eqn{rcMo2}, we arrive at
\beq \mathcal{M}_o = \f{4(\lp/2)^4}{\bigp{\lp/2+1}\bigb{\bigp{\lp/2}^2 + 1}} .\label{Mo32}\seq
For $\lp=2$, the only viable HEP value (see \fig{nearg1.5}), we immediately see that $\rh_c \equiv r_c/(\lp r_o) = 1/2$, that is, $r_c/r_o = 1$ and $\mathcal{M}_o = 1$.   There are thus no transonic wind solutions for $\gamma = 3/2$.  Values of $\lp>2$ do, however, yield the locations of the second set of roots under examination.  Notice that by \eqn{Mo32}, these roots correspond to critical point solutions with $\mathcal{M}_o > 1$.

The existence of two roots to the combined equations \eqref{rcMo1} and \eqref{rcMo2} was known to Parker, as he mentioned them in his original published account using a polytropic EoS (Parker, 1960).  Parker's analysis is insightful, using only Descartes' rule of signs, so we reproduce it in our notation.  For $\gamma<3/2$, equations \eqref{rcMo1} and \eqref{rcMo2} can be manipulated to read
\beq 2\bigp{\f{1}{\gamma-1} - \lp}\rh_c^\f{5-3\gamma}{\gamma-1} - e_c \rh_c^{\f{2(3-2\gamma)}{\gamma-1}} + \f{\lp^4}{2^\f{\gamma+1}{\gamma-1}} = 0.\label{Descartes1}\seq
From the HEP bound for $\gamma<3/2$, we see that the first coefficient is always positive and so there are two sign changes, meaning there are always two critical points.\footnote{Of course, Descartes rule only applies to polynomials, so this analysis applies to the infinite number of $\gamma$ in the range $1<\gamma<3/2$ that lead to integer exponents for the two $\rh_c$ terms. }  For $\gamma>3/2$, the exponent of the second term in \eqn{Descartes1} is negative, so the appropriate equation is now
\beq 2\bigp{\f{1}{\gamma-1} - \lp}\rh_c +\f{\lp^4}{2^\f{\gamma+1}{\gamma-1}}\rh_c^{\f{2(2\gamma-3)}{\gamma-1}} - e_c= 0.\label{Descartes2}\seq
In this case, the factor $2(2\gamma-3)/(\gamma-1)$ is never an integer in the range $3/2<\gamma<5/3$, so this line of analysis will not work.

\begin{figure*}
\centering
\includegraphics[scale=0.5]{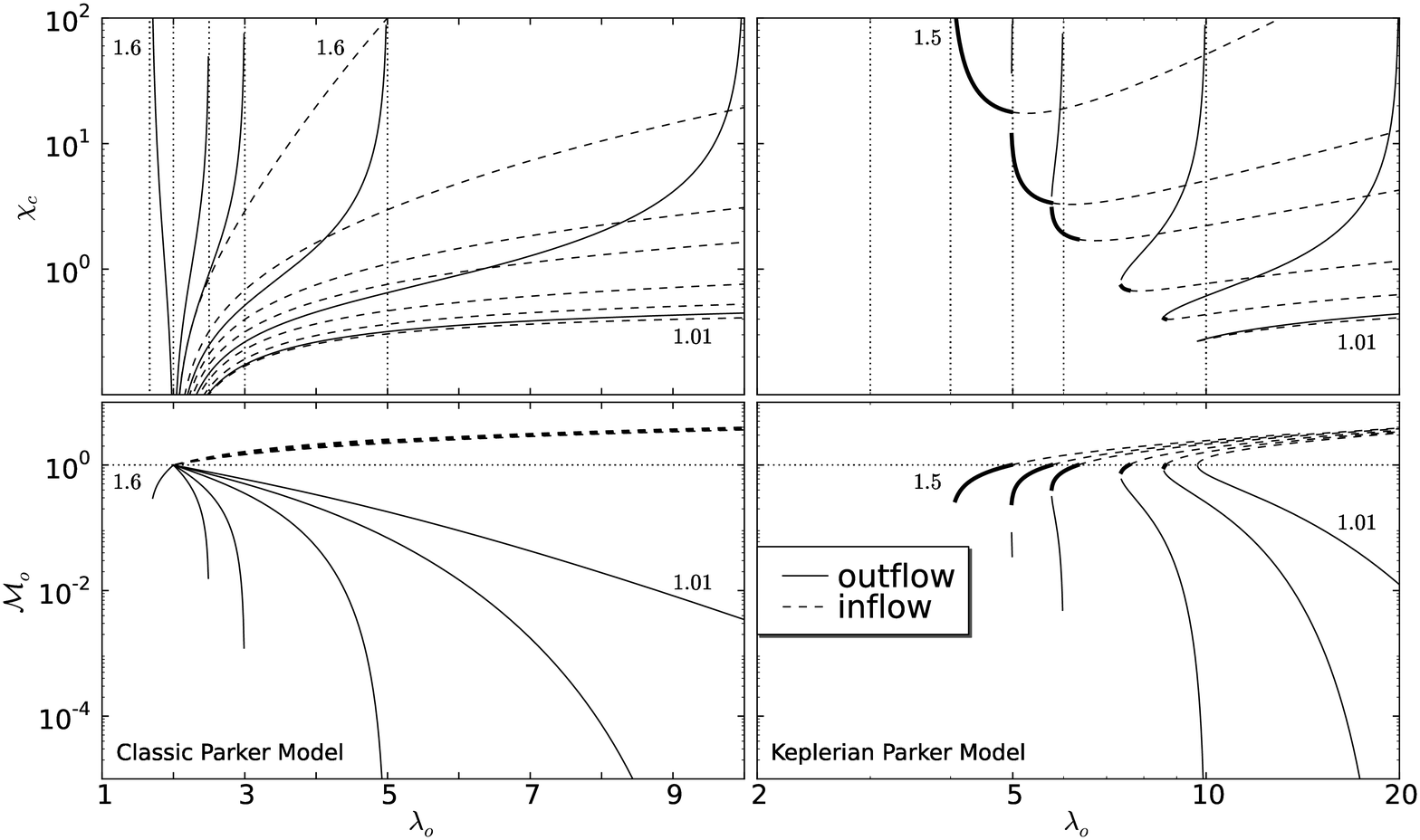}
\caption{Parameter survey for the Parker problem with and without Keplerian rotation.  The outflow critical point curves in between $\gamma = 1.01$ and $\gamma=1.6$ in the left panels are $\gamma = (1.1, 1.2,4/3,\&\,1.4)$; the inflow roots also have a $\gamma=1.5$ critical point curve.  The right panels, which have the same scaling as \fig{dwcritpts}, also have the intermediate $\gamma = (1.1, 1.2,4/3,\&\,1.4)$, but there are no solutions within $10^2\,r_g$ for $\gamma=1.6$.  Neither model has solutions for $\gamma=5/3$.  The vertical lines in the top left panel mark the values $1/(\gamma-1)$ (recall Table 1), whereas they denote the values $2/(\gamma-1)$ in top right panel.  Bold portions of curves have the same meaning as in \fig{dwcritpts}.  All plots in the main text display only outflow critical points.} 
\label{parkepsurvey}
\end{figure*}

\subsection{Properties of the `Inflow' Solutions}\label{inflowroots}
The second class of transonic solutions has two interpretations, one of them unphysical and the other physically acceptable but very unrealistic.  The latter case corresponds to a second outflow solution, in which the flow starts out supersonic and reaches a subsonic terminal velocity.  The former possibility is that of a transonic \textit{inflow} solution obeying \textit{inner} boundary conditions.  This is clearly physically unacceptable because transonic flows are insensitive to conditions downstream of the critical point.  Nevertheless, we choose to interpret these points as inflow solutions for the sake of classification and comparison.  With that choice, subsonic flow resides at $\x$ \textit{more distant} than the inflow critical point.

To illustrate the behavior of these inflow roots and how they can transition to outflow roots when rotation is added to the problem, we survey the parameter space of both the spherically symmetric and Keplerian Parker wind models.  \fig{parkepsurvey} is analogous to \fig{dwcritpts}, except that dashed critical point curves depict inflow roots.  The dashed curves in the top left panel are clearly of a different nature; they do not terminate at the vertical lines marking the values $1/(\gamma-1)$, meaning that the sonic point can reside well past $10^2 r_g$, as in the Bondi problem.  The bottom left panel shows that they all have approximately the same value of $\mathcal{M}_o$ (the terminal Mach number in this case).  A low HEP for the inflow solutions is interpreted as establishing a large back pressure which can prevent the flow from becoming sonic until it is very close to the star.  Notice that the inflow roots tend to the wind roots as $\gamma \rightarrow 1$ and degenerate into one root in the strictly isothermal case in which the critical points are all located at $\rh_c = \x_c + 1/\lp = 1/2$.  Finally, note that there is no regime change around $\gamma = 3/2$ for the inflow roots.  Again, for $\gamma = 3/2$, $\x_c$ vs. HEP and $\mathcal{M}_o$ vs. HEP are a priori known and given by equations \eqref{rc32} and \eqref{Mo32}, respectively.  

The inflow roots undergo a marked change in behavior upon adding Keplerian rotation to the Parker problem, as shown in the right panels.   The inflow and outflow curves become continuously connected, thereby accounting for the appearance of a tail.  The bolded portions of the outflow critical point curves would have stayed inflow curves had the density boundary condition $\rho(\x=0)/\rho_o=1$ remained satisfied by the transonic inflow solutions.  However, an inflow transonic solution would have to traverse \emph{two} sonic points in order to have a terminal Mach number less than unity.  Since this is prohibited, a inflow root becomes an outflow root once $\mathcal{M}_o < 1$.  In other words, degenerate wind solutions arose based on a mathematical requirement, which begs a time-dependent solution to the problem.  This occurrence may even be a further indication that inflow and outflow solutions are intimately coupled in such a way that the starting conditions of an accretion flow can lead to the subsequent onset of a wind (e.g., Blandford \& Begelman 1998).  

\end{document}